\RequirePackage{lineno}
\newif\ifpreprint%
\preprintfalse%
\ifpreprint%
	\documentclass[preprint,english,aps,prb,floatfix,amssymb,a4paper,10pt]{revtex4-2}
    
\else%
	\documentclass[twocolumn,english,aps,prb,floatfix,amssymb,a4paper,10pt]{revtex4-2}
\fi%
\usepackage{amsmath}
\usepackage{physics}
\usepackage{chemformula}
\usepackage{dsfont}
\usepackage{color}
\usepackage{amssymb}
\usepackage{bbold}
\usepackage{graphicx}
\usepackage{braket}
\usepackage{units}
\usepackage{lipsum}
\usepackage{hyperref}
\hypersetup{colorlinks = true, linkcolor=magenta,citecolor=blue, urlcolor=magenta, bookmarksnumbered =  true}

\newcommand{\ssm}{\scriptscriptstyle\rm}

\renewcommand{\theta}{\vartheta}
\renewcommand{\phi}{\varphi}

\renewcommand{\vec}[1]{\boldsymbol{#1}}

\newcommand{\crea}[1]{#1^{\dag}}
\newcommand{\anni}[1]{#1^{\vphantom{\dag}}}
\newcommand{\up}{\uparrow}
\newcommand{\down}{\downarrow}

\renewcommand{\braket}[2]{\langle#1|#2\rangle}

\begin{document}
\ifpreprint%
	\linenumbers%
\fi%

\title{Superfluid stiffness of superconductors with delicate topology}

\author{
Tijan Prijon}
\author{
Sebastian D. Huber}
\author{
Kukka-Emilia Huhtinen}
\email{khuhtinen@phys.ethz.ch}
\affiliation{
Institute for Theoretical Physics, ETH Zurich, 8093 Zürich, Switzerland
}

\begin{abstract}
    We consider superconductivity in two-dimensional delicate topological bands, where the total Chern number vanishes but the Brillouin zone can be divided into subregions with a quantized nontrivial Chern number. We formulate a lower bound on the geometric contribution to the superfluid weight in terms of the sum of the absolute values of these sub-Brillouin zone Chern numbers. We verify this bound in Chern dartboard insulators, where the delicate topology is protected by mirror symmetry. In iso-orbital models, where the mirror representation is the same along all high-symmetry lines, the lower bound increases linearly with the number of mirror planes. This work points to delicate bands as promising candidates for particularly stable superconductivity, especially in narrow bands where the kinetic energy is suppressed due to lattice effects.
\end{abstract}

\date{\today}

\maketitle

\section{Introduction}

With the advent of two-dimensional materials, an exciting playground for unconventional superconductivity opened up. In particular, superconductivity seems to be stabilized in a variety of materials where the electronic bands are characterized by a small bandwidth~\cite{Mielke2021,Khasanov2024,Cao2018,Lu2019,Hao2021,Park2021,Park2022,Zhou2021b} which renders them inherently strongly correlated. This re-opened the question on what stabilizes such quantum coherent states in systems where the free electrons are strongly localized. 

Stable superconductivity requires a non-singular superfluid weight tensor, or phase stiffness, $D$, as it controls both the Meissner effect and dissipationless transport \cite{Schrieffer1964,Scalapino1993}. Moreover, in two-dimensional materials the superfluid weight determines the critical temperature $T_{\rm BKT} = \frac{\pi}{8}\sqrt{{\rm det}[D]}$ through the Berezinskii-Kosterlitz-Thouless (BKT) mechanism \cite{Berezinskii1970,Kosterlitz72,Kosterlitz73,Nelson77}.

In a single band, the superfluid weight follows the text book result $D\propto  n_s/m^*$, where $n_s$ is the density of paired electrons and $m^*$ is the effective mass tensor. The effective mass is determined by the details of the band dispersion, and in particular, a flat dispersion corresponding to an infinite effective mass leads to a vanishing superfluid weight.

In a multi-band lattice, however, this picture is not complete. Instead, the superfluid weight contains a contribution which depends on the geometry of the Bloch states. This geometric contribution is generally present in any multi-band model, and is especially important in the narrow and flat-band limit.

In a perfectly flat isolated band, the superfluid weight is proportional to the minimal quantum metric~\cite{Peotta2015,Huhtinen2022,Peotta2024}. The relationship between the quantum metric and flat-band superconductivity is not only a mean-field effect, but is consistent with results from numerical methods such as quantum Monte Carlo~\cite{Peri2021,Herzog-Arbeitman2022}, dynamical mean-field theory~\cite{Liang2017,Penttila2025}, DMRG~\cite{Tovmasyan18a} and exact diagonalization~\cite{Julku2016,Liang2017}. It can also be proved exactly under certain conditions~\cite{Tovmasyan16,Herzog-Arbeitman2022b}, and  manifests already in the two-body effective mass~\cite{Torma2018,Iskin2021,Iskin2022}. The quantum metric $\mathcal{M}_{\mu\nu}^{\mathcal B}({\boldsymbol k})$ is the real part of the quantum geometric tensor (QGT)
\begin{equation}
\begin{aligned}
    \mathcal{G}^{\mathcal{B}}_{\mu\nu}(\vec{k}) \equiv& 2\sum_{m\in\mathcal{B}} \sum_{n\notin\mathcal{B}} \braket{\partial_{\mu}m_{\vec{k}}}{n_{\vec{k}}} \braket{n_{\vec{k}}}{\partial_{\nu}m_{\vec{k}}} \\
    =& \mathcal{M}^{\mathcal{B}}_{\mu\nu}(\vec{k}) - i\mathcal{F}^{\mathcal{B}}_{\mu\nu}(\vec{k}),
\end{aligned}
\label{eq.qgt}
\end{equation}
where $\mathcal{B}$ denotes a set of bands, $|m_{\boldsymbol k}\rangle$ is the Bloch eigenstate of band $m$ and $\partial_\mu$ is the derivative with respect to the $\mu$th component of the momentum $\boldsymbol k$. The imaginary part of the quantum geometric tensor gives the well known Berry curvature $\mathcal{F}^{\mathcal{B}}_{\mu\nu}(\vec{k})$. The fact that the quantum geometric tensor $\mathcal{G}^{\mathcal{B}}_{\mu\nu}(\vec{k})$ is positive semidefinite \cite{Peotta2015,Roy2014} makes it possible to establish relations between its real and imaginary parts and hence one can formulate bounds on the metric which relate to the band topology encoded in the Berry curvature.

Several lower bounds on the quantum metric have been established in terms of Chern numbers~\cite{Roy2014,Peotta2015,Ozawa2021,Mera2022}, winding numbers~\cite{Tovmasyan16}, the Euler class~\cite{Xie2020,Jankowski2024}, real-space invariants~\cite{Herzog-Arbeitman2022}, and most recently $\mathbb Z_2$ invariants and more generally the absolute winding of Wilson loops~\cite{Yu2025}. Typically, these results are used to formulate bounds on the trace of the superfluid weight ${\rm Tr}[D]$. In isotropic systems~\footnote{The superfluid weight tensor is of the form $D\propto\mathbb{1}$ when the model is sufficiently symmetric, with for instance $C_3$ or $C_4$ rotational symmetry. We refer to these models as isotropic.}, $D$ is proportional to the identity, and considering the trace is sufficient. However, in anisotropic models, ${\rm Tr}[D]> 0$ does not guarantee ${\rm det}[D]> 0$, i.e., the trace alone is not sufficient to predict stability.

Here, we focus on a new class of topological bands, dubbed delicate topological~\cite{Nelson2021,Nelson2022,Moore2008,Liu2017}. The name delicate comes from the fact that the band can be trivialized by adding a band {\em above} the gap. This is in striking contrast to strong or fragile topology, where one has to add another topological or trivial band, respectively, {\em below} the gap to trivialize the band~\cite{Bradlyn17, Po2018}. This renders delicate topology a property of the {\em full} Hilbert space, not just the {\em filled} bands.

In particular, we consider Chern dartboard insulators (CDIs)~\cite{Chen2024}. These are systems with a vanishing Chern number, where, however, mirror symmetries can lead to non-trivial and quantized  sub-Brillouin zone Chern numbers $C_i$. These delicate topological systems can be separated into two distinct classes, depending on whether the mirror representation is the same at all high-symmetry lines, which we dub iso-orbital, or different, i.e. aniso-orbital. As we show below, this distinction is crucial when it comes to superconductivity: only in the former case do the sub-Brillouin zone Chern numbers have an impact on the superfluid weight. The main result of our work is that in the formulation of bounds on $\det[D]$, the sub-Brillouin zone Chern numbers $C_i$ of iso-orbital models add up as $\sum_i |C_i|$, yielding lower bounds on the superfluid stiffness that exceed those of previously known examples. The bound we formulate is valid in both isotropic and anisotropic models, and can give a lower estimate for the importance of interband processes also in dispersive bands. Our results show that delicate bands are a particularly promising platform for remarkably stable flat-band superconductivity due to their highly nontrivial quantum geometry.

\section{Sub-Brillouin zone topology}

We work with a system with $M$ orbitals and assume that the set $\mathcal B$ of $n_b$ bands of interest has a trivial Chern number. Moreover, we require that the total Brillouin zone (BZ) $\Omega$ can be divided into regions $\Omega_i$, where along the boundaries $\partial \Omega_i$ the $n_b$ bands occupy a fixed set of exactly $n_b$ orbitals. Let us define the $M \cross n_b$ matrix $\mathcal{U}_{\vec{k}}$ such that the $m$th column is the Bloch state $\ket{m_{\vec{k}}}$ of the $m$th band in $\mathcal{B}$. For now, we work with a Fourier transformation convention that makes the single-particle Hamiltonian periodic, $H_{\vec{k}}=H_{\vec{k}+\vec{G}}$, where $\vec{G}$ is a reciprocal lattice vector. We refer to the choice of Fourier transformation convention, i.e. to the choice of orbital positions, as a basis. Since the total Chern number is zero, the Bloch states can be chosen to be smooth and periodic $\ket{m_{\vec{k}+\vec{G}}} = \ket{m_{\vec{k}}}$. The Berry curvature of the set of bands $\mathcal{B}$ is $\mathcal{F}_{xy}^{\mathcal{B}}(\vec{k}) = i[\nabla\times{\rm Tr}(\mathcal{U}_{\vec{k}}^{\dag}\nabla  \mathcal{U}_{\vec{k}}^{\vphantom{\dag}})]_z$. The Chern number over the region $\Omega_i$ is given by
\begin{equation}
  \begin{aligned}
    C_i =& \frac{1}{2\pi}\int_{\Omega_i} d\vec{k}^2
    \mathcal{F}_{xy}^{\mathcal{B}}(\vec{k}) \\
    =& \frac{i}{2\pi}\oint_{\partial\Omega_i} d\vec{k}\cdot
    {\rm Tr}[\mathcal{U}_{\vec{k}}^{\dag}\nabla
    \mathcal{U}_{\vec{k}}^{\vphantom{\dag}}]. \label{eq.line_int}
  \end{aligned}
\end{equation}

Since we assume that the $n_b$ bands in $\mathcal B$ occupy $n_b$ fixed orbitals at the boundary $\partial\Omega_i$, all the non-vanishing elements of $\mathcal{U}_{\vec{k}}$ are on $n_b$ rows (for $n_b$ orbitals) when $\vec{k}$ is on the boundary. We can thus discard the rows with only vanishing elements at $\partial\Omega_i$ to obtain a reduced $n_b \times n_b$ square matrix denoted $\mathcal{V}_{\vec{k}}$ for which ${\rm Tr}[\mathcal{U}_{\vec{k}}^{\dag}\nabla\mathcal{U}_{\vec{k}}]={\rm Tr}[\mathcal{V}_{\vec{k}}^{\dag}\nabla\mathcal{V}_{\vec{k}}]$ for any $\vec{k}\in\partial\Omega_i$. The expression for the Chern number now reads
\begin{equation}
  C_i = \frac{i}{2\pi}\oint_{\partial\Omega_i} d\vec{k}\cdot
    {\rm Tr}[\mathcal{V}_{\vec{k}}^{\dagger}\nabla
    \mathcal{V}_{\vec{k}}^{\vphantom{\dagger}}].
\end{equation}
$\mathcal{V}_{\vec{k}}$ defines a pure ${\rm U}(n_b)$ gauge and hence $C_i$ is indeed quantized. 

For the considered CDIs~\cite{Chen2024}, the existence of regions in the BZ fulfilling the above assumptions is due to mirror symmetries: if along the high-symmetry lines (HSLs) in the BZ all $n_b$ bands in $\mathcal{B}$ have the same mirror eigenvalue, and, importantly, none of the other bands is in the same mirror representation, the Chern number in each irreducible BZ $\Omega_i$ is well defined and quantized. 

\begin{figure}
    \centering
    \includegraphics[width=\columnwidth]{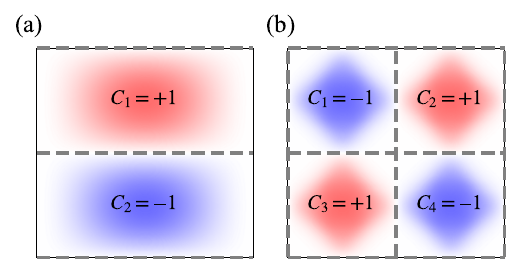}
    \caption{Berry curvature $\mathcal{F}^{\mathcal{B}}_{xy}$ in the first Brillouin zone for (a) the model with one mirror symmetry, and (b) the model with two mirrors (see supplemental material~\cite{supplemental} for Hamiltonians). Positive values of $\mathcal{F}^{\mathcal{B}}_{xy}$ are shown in red, and negative in blue. Dashed lines indicate mirror high-symmetry lines that partition the Brillouin zone into subregions with well defined quantized Chern numbers.}
    \label{fig:berry_curv}
\end{figure}

One subtlety remains regarding the quantization of the Chern numbers. If we consider more than one mirror plane, all $\Omega_i$ are fully bounded by HSLs, cf. Fig.~\ref{fig:berry_curv}.b and the arguments above directly apply. However, for the case of one mirror, the BZ boundaries are part of $\partial \Omega_i$, cf. Fig.~\ref{fig:berry_curv}.a. In a periodic basis, these two boundaries can be identified, and the Chern numbers over the irreducible BZs are given by
\begin{equation}
  C_1 = \frac{i}{2\pi}\int d\vec{k}\cdot {\rm Tr}[\mathcal{V}_{\pi}^{\dag}\nabla
  \mathcal{V}_{\pi} - \mathcal{V}_0^{\dag}\nabla \mathcal{V}_0],
\end{equation}
and $C_2=-C_1$, with $\mathcal{V}_{0}$ and $\mathcal{V}_{\pi}$ the matrix $\mathcal{V}_{\vec{k}}$ at the two HSLs in $\partial \Omega_i$. $C_1$ is the difference of two integers and therefore an integer. However, this statement is only true in a periodic basis. As the superfluid weight has to be basis invariant, this point requires further attention ~\cite{Huhtinen2022,Peotta2024}. It is easy to see \cite{Nelson2022, Chen2024}, that if along both HSLs, the same orbitals are occupied (iso-orbital), the Chern number is the same in any basis. In the opposite case, where along the two HSLs the mirror representation is opposite (aniso-orbital), the numbers $C_i$ can take any value and are only quantized in specific bases~\cite{supplemental}.

\section{Lower bound on the superfluid weight}

We now derive a mean-field bound for the superfluid weight in terms of the sub-BZ Chern numbers $C_i$. The mean-field Hamiltonian can be written as $H_{\ssm MF} = \sum_{\vec{k}}\crea{\psi}_{\vec{k}}H_{\rm \scriptscriptstyle  BdG}(\vec{k})\anni{\psi}_{\vec{k}}$, where $\psi_{\vec{k}}^{\dag} =(\crea{c_{\vec{k}(\alpha=1)\up}},\ldots,\crea{c_{\vec{k}(\alpha=M)\up}},\anni{c_{-\vec{k}(\alpha=1)\up}},\ldots,\anni{c_{-\vec{k}(\alpha=M)\up}})^T$ and
\begin{equation}
  H_{\rm \scriptscriptstyle  BdG}(\vec{k}) = \begin{pmatrix}
    H_{\vec{k}}^{\up} - \mu\mathbb{1} & \vec{\Delta} \\
    \vec{\Delta}^{\dag} & -(H_{-\vec{k}}^{\down})^*+\mu\mathbb{1}
  \end{pmatrix},
\end{equation}
$H_{\vec{k}}^{\sigma}$ denotes the kinetic Hamiltonian for spin $\sigma$ and $\vec{\Delta} = {\rm diag}(\Delta_1,\ldots,\Delta_M)$, with $\Delta_{\alpha} = (U/N)\sum_{\vec{k}} \langle c_{\vec{k}\alpha}c_{-\vec{k}\alpha}\rangle$ being the mean-field order parameters. We consider systems with time-reversal symmetry (TRS) such that $H_{\vec{k}}\equiv H_{\vec{k}}^{\up}=(H_{-\vec{k}}^{\down})^*$.

The superfluid stiffness can be defined as the $\omega=0$, $\vec{q}\to 0$ limit of the current-current response function $K_{\mu\nu}$ in $j_{\mu}(\vec{q},\omega) =
K_{\mu\nu}(\vec{q},\omega)A_{\nu}(\vec{q},\omega)$, where $\vec{j}(\vec{q},\omega)$ is the current density and $A_{\nu}(\vec{q},\omega)$ is the vector potential. Note that in mean-field theory the order parameters $\Delta_{\alpha}$ can also depend on $\vec{A}$, which generally needs to be taken into account when computing the superfluid weight~\cite{Huhtinen2022}. Since we assume TRS, there exists a basis where $\delta\Delta_{\alpha}/\delta A_{\mu}=0$ in the $\vec{q}\to 0$ limit. This is the basis we work in to derive the bound. Note that the basis does not need to be periodic. As a consequence, in aniso-orbital models, the Chern number might not be quantized in the particular basis we work in, but this does not impact the arguments below. 

The superfluid weight in this basis is given by 
~\cite{Liang2017,Huhtinen2022}
\begin{equation}
  \begin{aligned}
    D_{\mu\nu} =& \frac{1}{N}\sum_{\vec{k},i,j} 
    \frac{n_F(E_j)-n_F(E_i)}{E_i-E_j}\\
    &\bigg(
    \bra{\psi_i}\partial_{\mu}H_{\rm \scriptscriptstyle  BdG}(\vec{k}) \ket{\psi_j}
    \bra{\psi_j}\partial_{\nu}H_{\rm \scriptscriptstyle  BdG}(\vec{k})\ket{\psi_i}  \\
    &- \bra{\psi_i}\partial_{\mu} H_{\rm \scriptscriptstyle  BdG}(\vec{k})\gamma^z
    \ket{\psi_j} \bra{\psi_j} \partial_{\nu}H_{\rm \scriptscriptstyle  BdG}(\vec{k})\gamma^z
    \ket{\psi_i} \bigg).
  \end{aligned}
   \label{eq.dsmt}
\end{equation}
Here, $E_i$ and $\ket{\psi_i}$ are the eigenvalues and eigenvectors of $H_{\rm \scriptscriptstyle  BdG}$, respectively, and $\gamma^z = \sigma_z\otimes \mathbb{1}_M$, with $\mathbb{1}_M$ the $M\times M$ identity matrix. Moreover, $n_F(E) = 1/(e^{\beta E}+1)$ is the Fermi-Dirac distribution and $\beta$ is the inverse temperature. When $E=E_i=E_j$, the prefactor should be understood as $-\partial_En_F(E)$. Note that since we assume $\vec{\Delta}$ to be independent of $\vec{k}$, the $\vec{k}$-derivatives $\partial_{\mu}\equiv \partial/\partial k_{\mu}$ of $H_{\rm \scriptscriptstyle  BdG}$ contain only derivatives of the kinetic Hamiltonian.

By expressing the eigenvectors of $H_{\rm \scriptscriptstyle  BdG}$ in terms of Bloch functions, it is possible to split Eq.~\eqref{eq.dsmt} into a conventional and geometric part, $D=D^{\rm conv} + D^{\rm geom}$, where $D^{\rm conv}$ contains only intraband terms, while $D^{\rm geom}$ involves interband terms. The diagonalization of $H_{\rm\scriptscriptstyle BdG}$ often needs to be performed numerically, but can be done analytically if we assume the uniform pairing conditions (UPC), $\Delta_{\alpha}=\Delta$. Expressions for $D^{\rm conv}$ and $D^{\rm geom}$ under this assumption are given in the supplemental material~\cite{supplemental}. The
conventional contribution depends on the derivatives of the band
dispersions, and does not involve quantum geometry. The geometric contribution, on the other hand, involves terms of the form $\braket{\partial_{\mu}m_{\vec{k}}}{n_{\vec{k}}}\braket{n_{\vec{k}}}{\partial_{\nu}m_{\vec{k}}}$. These terms also appear in the quantum metric.

The QGT \eqref{eq.qgt} is positive semidefinite at every momentum, which implies that ${\rm det}[\mathcal{G}^{\mathcal{B}}(\vec{k})] = {\rm det}[\mathcal{M}^{\mathcal{B}}(\vec{k})] - |\mathcal{F}^{\mathcal{B}}_{xy}(\vec{k})|^2 \geq 0$ at every $\vec{k}$. This equation allows for the formulation of topological bounds on $D^{\rm geom}$. 
However, we need $\mathcal{M}^{\mathcal{B}}$, to appear, not just $\braket{\partial_{\mu}m_{\vec{k}}}{n_{\vec{k}}}\braket{n_{\vec{k}}}{\partial_{\nu}m_{\vec{k}}}$ weighted by different factors. This is not the case in general, but if
we assume a set of degenerate bands $\mathcal{B}$ at the Fermi level 
is well-isolated from the other bands,
i.e. 
$\varepsilon_n=\varepsilon_{\overline{n}}$, 
$|\varepsilon_n-\varepsilon_m|\gg |\varepsilon_n-\mu|$ and
$|\varepsilon_n-\varepsilon_m|\gg|\Delta|$ 
 for all $n\in\mathcal{B}$  
and $m\notin\mathcal{B}$, $D^{\rm geom}$ can be approximated in the thermodynamic limit by
\begin{equation}
  D^{\rm geom}_{\mu\nu} = |\Delta^2|\int_{\rm B.z.} d\vec{k}^2
  f^{\mathcal{B}}(\vec{k}) \mathcal{M}^{\mathcal{B}}(\vec{k}), \label{eq.ds_geom_isolated}
\end{equation}
where $f^{\mathcal{B}}(\vec{k}) = 2\tanh(\beta
E_{\overline{n}}(\vec{k})/2)/E_{\overline{n}}(\vec{k})/(2\pi)^2$ is a positive function.

This result holds for both flat and dispersive bands under the
assumption of TRS, isolated bands and uniform pairing. In flat bands,
UPC is equivalent to $(1/2\pi)^2\int_{\rm B.z.}d\vec{k} ^2[P_{\mathcal{B}}]_{\alpha\alpha} =
1/n_{\rm orb}$~\cite{Tovmasyan16}, where $P_{\mathcal{B}} =
\sum_{n\in\mathcal{B}}\ket{n}\bra{n}$ and $n_{\rm orb}$ is the
number of orbitals occupied by the states in $\mathcal{B}$,
i.e. UPC can be verified from the properties of the non-interacting
Hamiltonian. In dispersive bands, however, these two conditions are no
longer equivalent, and how well UPC is fulfilled typically depends on
the interaction. 

We now divide the BZ into regions $\Omega_i$, and aim to
establish a lower bound on ${\rm det}[D]$ in terms of the integral of
the Berry curvature over these regions,
\begin{equation}
  C_{i} = \frac{1}{2\pi}\int_{\Omega_i}d\vec{k}^2 \mathcal{F}^{\mathcal{B}}_{xy}(\vec{k}). 
\end{equation}
Note that while we denote this integral as $C_{i}$, the
arguments in the following do not require $C_{i}$ to be
quantized. However, when applying our result to iso-orbital Chern dartboard
insulators, $C_i$ will be a well-defined Chern number. 

Both $D^{\rm conv}$ and $D^{\rm geom}$ are positive semidefinite under our assumptions, and therefore ${\rm det}[D]\geq {\rm det}[D^{\rm conv}]+{\rm det}[D^{\rm geom}]$. Any lower bound on ${\rm det}[D^{\rm geom}]$ is thus also a lower bound for ${\rm det}[D]$. 

Minkowski's determinant inequality for positive semidefinite $2\times 2$ matrices implies that
\begin{equation}
    {\rm det}[D^{\rm geom}] \geq |\Delta|^4\left( \int_{\rm B.z.}d\vec{k}^2 \sqrt{ {\rm
        det}[f^{\mathcal{B}}(\vec{k})
        \mathcal{M}^{\mathcal{B}}(\vec{k})] } \right)^2
\end{equation}
When $\Delta\neq 0$, the minimum ${\rm min}(f^{\mathcal{B}})$ over the BZ is strictly positive. We take $f^{\mathcal{B}}$ out of the momentum integral by using that $f^{\mathcal{B}}\geq {\rm min}(f^{\mathcal{B}})>0$ and that ${\rm det}[\mathcal{M}^{\mathcal{B}}(\vec{k})]\geq 0$ at all $\vec{k}$ to obtain 
\begin{equation}
  \begin{aligned}
    {\rm det}[D] \geq |\Delta|^4({\rm
      min}(f^{\mathcal{B}}))^2
    \bigg[\int_{\rm B.z.} d\vec{k}^2 \sqrt{{\rm det}[\mathcal{M}^{\mathcal{B}}(\vec{k})]}\bigg]^2 
  \end{aligned}
\end{equation}
Because the quantum geometric tensor is positive semidefinite at all
$\vec{k}$, we have that
\begin{equation}
  \begin{aligned}
    {\rm det}[D] \geq {\rm det}[D^{\rm geom}] \geq& |\Delta|^4 ({\rm
      min}(f^{\mathcal{B}}))^2
    \bigg[\int_{\rm B.z.}d\vec{k}^2|\mathcal{F}^{\mathcal{B}}_{xy}(\vec{k})|\bigg]^2 \\
    \geq& (2\pi)^2|\Delta|^4 \left({\rm
      min}(f^{\mathcal{B}})\right)^2 \bigg[\sum_i |C_{i}|\bigg]^2.
  \end{aligned}
\end{equation}
This result holds only in an appropriate basis, where the trace of
$\sum_{\vec{k}}\mathcal{M}^{\mathcal{B}}_{\mu\nu}$ is
minimized. However, if $|C_{i}|$ are well-defined Chern numbers
that are basis-invariant, the bound itself is basis-invariant and thus
holds in any basis. This is the case for the delicate topological bands
considered here when the iso-orbitality condition is fulfilled, but
not in the aniso-orbital model. 

In the particular case of the iso-orbital Chern dartboard insulators, $\sum_i|C_{i}| = 2n_{\rm M}|C|$, where $C$ is the Chern number in the irreducible BZ and $n_{\rm M}$ is the number of mirrors. Then
\begin{equation}
  \sqrt{{\rm det}[D]} \geq 2\pi|\Delta|^2n_{\rm M}{\rm min}(f^{\mathcal{B}})|C|.
  \label{eq:final_bound}
\end{equation}
giving a lower bound on the superfluid weight that grows linearly with the number of mirrors.

\section{Numerical results}

To verify that the bound holds, we numerically compute the superfluid weight in CDI models fulfilling the iso-orbital condition and with $n_{\rm M}=1,2,3,4$ mirrors presented in~\cite{Chen2024}. We first focus on these iso-orbital models since they have a well-defined basis-independent Chern number in the irreducible BZ, cf. Fig.~\ref{fig:berry_curv}. 

The non-interacting Hamiltonians can be written as $H^{n_{\rm M}}_{\vec{k},m} = \vec{\sigma} \cdot\vec{d}^{n_{\rm M}}(\vec{k},m)$, where $\vec{\sigma}=(\sigma_x,\sigma_y,\sigma_z)$ is the vector of Pauli matrices. The corresponding $\vec{d}^{n_M}(\vec{k},m)$ vectors for the models with $n_{\rm M}$ mirror symmetries are given in the supplemental material~\cite{supplemental}. The on-site energy difference of the two orbitals is tuned by $m$. For $0<m<2$, the models are gapped and have a nonzero quantized Chern number, $|C|=1$, in the irreducible BZ. At $m=0,2$, band touchings occur, inducing a topological transition to trivial bands.

We first focus on the flat-band limit, where the geometric contribution is dominant. Only the single mirror model can be tuned to a flat band at $m=1$, while the other models remain dispersive for all values of $m$. To investigate how the superfluid weight behaves in the flat-band limit also for larger $n_{\rm M}$, we construct flat-band models with the same Bloch functions as the models given in the supplemental material~\cite{supplemental}.

To achieve this, we first diagonalize the kinetic Hamiltonian as $H_{\vec{k},m}^{n_{\rm M}} = \mathcal{G}_{\vec{k},m}\boldsymbol{\varepsilon}_{\vec{k},m}\mathcal{G}^{\dag}_{\vec{k},m}$. We then replace the matrix containing the dispersions $\boldsymbol{\varepsilon}_{\vec{k},m}$ by  $\sigma_z$ to obtain the model $H_{\vec{k},m}^{n_{\rm M},{\rm FB}} = \mathcal{G}_{\vec{k},m}\sigma_z\mathcal{G}^{\dag}_{\vec{k},m}$ which has flat bands at $E=\pm 1$ and the same Bloch functions as the original dispersive bands. The drawback of this method is that the hopping amplitudes generically become long-ranged. However, this allows us to obtain models with non-dispersive bands, so $D^{\rm conv}=0$ and the superfluid weight is fully geometric.

\begin{figure}
    \centering
    \includegraphics[width=\columnwidth]{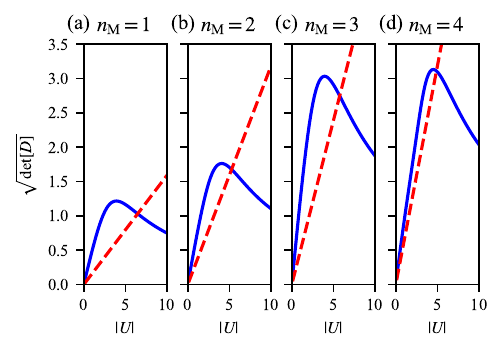}
    \caption{Zero-temperature superfluid weight (blue) in flat-band models with (a) $n_{\rm M}=1$, (b) $n_{\rm M}=2$, (c) $n_{\rm M}=3$ and (d) $n_{\rm M}=4$ mirrors, as a function of the interaction $U$. The red dashed lines indicate the bound derived in terms of the Chern number in the irreducible Brillouin zone [Eq.~\eqref{eq:flat_band_bound}]. The slope of the bound grows linearly with $n_{\rm M}$. The bound is derived for an isolated flat band with uniform pairing, and holds for low interactions. The pairing is not perfectly uniform for all models, but the bound still holds. The difference in on-site energies of the two orbitals is set to  $m=0.25$, a value at which the pairing is almost uniform for $n_{\rm M}=1$ and $n_{\rm M}=3$ (see Fig.~\ref{fig:flatband_param}). The lowest flat band is half-filled.}
    \label{fig:fb_d_vs_u}
\end{figure}

For a single perfectly flat isolated band with UPC, the zero-temperature
$\vec{\Delta}$  and $f^{\mathcal{B}}$ can be solved analytically, and Eq.~\eqref{eq:final_bound} can be expressed in terms of the interaction $U$ and filling fraction $\nu$ as
\begin{equation}
  \sqrt{{\rm det}[D]} \geq \frac{2}{\pi}|U|\nu(1-\nu)n_{\rm M}|C|.
  \label{eq:flat_band_bound}
\end{equation}
This bound is plotted as a function of interaction along with the numerically obtained superfluid weight in Fig.~\ref{fig:fb_d_vs_u} at $m=0.25$. The bound indeed holds for this value of $m$ at low interactions, i.e. where the isolated band assumption holds, for all four models considered.

\begin{figure}
    \centering
    \includegraphics[width=\columnwidth]{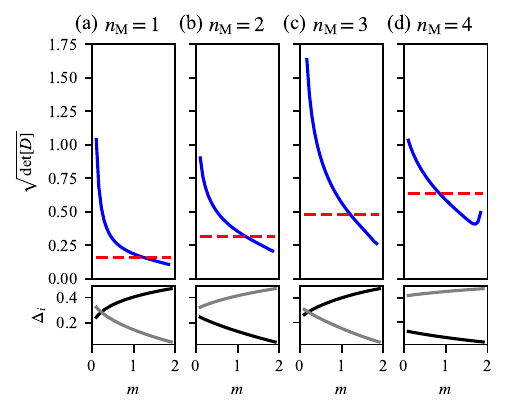}
    \caption{Upper panels: Superfluid weight (blue) for flat-band models with (a) $n_{\rm M}=1$, (b) $n_{\rm M}=2$, (c) $n_{\rm M}=3$ and (d) $n_{\rm M}=4$ mirror symmetries, together with the bound from Eq.~\eqref{eq:flat_band_bound} (red dashed line), as a function of $m$. At $m=0$ and $m=2$, the gap closes in the single-particle spectrum, and $|C|=1$ in the irreducible BZ in between these values. Lower panels: Order parameters $\Delta_1$ and $\Delta_2$. The UPC is fulfilled when $\vec{\Delta} \propto \mathbb{1}$, i.e. when the lines intersect, which occurs only for $n_{\rm M}=1,3$ close to $m\approx0.25$. In all panels, the filling fraction of the lowest band is set to $\nu=1/2$, and the interaction to $U=-1$.}
    \label{fig:flatband_param}
\end{figure}
However, UPC holds perfectly only for single values of $m$, only in the $n_{\rm M}=1$, and $n_{\rm M}=3$ models. At least in some models, it is possible to analytically obtain the superfluid weight even in the absence of UPC~\cite{Chan2022}, but these expressions typically no longer involve the quantum metric in a simple way. As can be seen from Fig.~\ref{fig:flatband_param} the bound Eq.~\eqref{eq:flat_band_bound}, which assumes UPC, remains robust well into the regime where the UPC no longer holds, and only breaks once one component of $\vec{\Delta}$ exceeds the other by roughly a factor of $2$. Note that although the bound grows linearly with $n_{\rm M}$, there is no a priori way to predict how tight the bound will be. It is therefore possible for the actual superfluid weight to behave non-monotonically with $n_{\rm M}$.

The bound Eq.~\eqref{eq:final_bound} should also hold when the band is not flat as long as it is isolated. We therefore test the bound on the geometric contribution to $D$ using dispersive models described in the supplemental material~\cite{supplemental}, without flattening the bands. In contrast to the models with only flat bands, there is now also a conventional contribution to the superfluid weight which we expect to be dominant. Moreover, $\Delta$, which appears in the prefactor of Eq.~\eqref{eq:final_bound}, is expected to be exponentially suppressed when $W\gg |U|$, where $W$ is the bandwidth. This is in contrast to the isolated flat-band limit where $\Delta$ grows linearly with $|U|$. The bound in terms of $|C|$ is a lower bound for the geometric contribution, and is thus expected to be quite loose for the full superfluid weight when bands are very dispersive, which is the case for our models. As $|U|$ becomes larger compared to the bandwidth, the system should get closer to flat-band behavior~\cite{Thumin2025}, with a larger geometric contribution and less dominant conventional part.

Unlike in the flat-band limit, multiband systems with dispersive bands admit no closed form expression relating $\vec{\Delta}$ to the filling fraction $\nu$ and interaction $U$ (even at $T=0$). Thus, outside of the perfect UPC regime one cannot unambiguously generalize the bound from Eq.~\eqref{eq:final_bound}, which requires a single value $\Delta$. However, to examine the behavior of the bound under (slight) UPC violations, we approximate $\vec{\Delta} \rightarrow ({\rm Tr}[\vec{\Delta}]/2)\mathbb{1}$. This should be regarded as an estimate rather than a strict bound.

\begin{figure}
    \centering
    \includegraphics[width=\columnwidth]{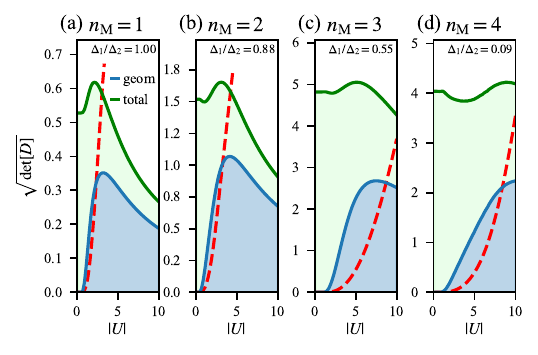}
    \caption{Superfluid weight computed for dispersive models with (a) $n_{\rm M}=1$, (b) $n_{\rm M}=2$, (c) $n_{\rm M}=3$ and (d) $n_{\rm M}=4$, together with the bound from  Eq.\eqref{eq:final_bound} (red dashed line). The blue line corresponds to the geometric part $D^{\rm geom}$, and the green line to the total superfluid weight including also the conventional contribution $D^{\rm conv}$. The ratio $\Delta_{1}/\Delta_2$ is evaluated in the $U\to0$ limit to evaluate how well the UPC is satisfied, $\Delta_1/\Delta_2=1$ indicating uniform pairing. The UPC is fulfilled only for $n_{\rm M}=1$, for other models the bound serves only as an approximation. The filling fraction is set to $\nu=1/2$, and $m=0.25$.}
    \label{fig:dispersive}
\end{figure}

With this approximation, as can be seen in Fig.~\ref{fig:dispersive}, the bound lies below numerical values of $\sqrt{{\rm det}[D^{\ssm geom}]}$ and also accurately captures the functional dependence of $\sqrt{{\rm det}[D^{\ssm geom}]}$ at low $U$ for $n_{\rm M}=1,2$. For these models, $\Delta_1/\Delta_2$ is close to one, and UPC is almost fulfilled. For $n_{\rm M}=3$ and $n_{\rm M}=4$, the bound still remains below the geometric contribution even though UPC is strongly broken. This is partly because the bands are more dispersive, leading to larger variations of $f^{\mathcal B}$, and thus a naturally looser bound. Overall, the bound is relatively loose for all models compared with the total superfluid weight, since it is lower estimate only of the geometric part. However, the bound gives a lower estimate for the magnitude of $D^{\rm geom}$, which in some cases including $n_{\rm M}=1$ and $n_{\rm M}=2$ is fairly close to the actual value, including at interactions where the geometric contribution is non-negligible.

\begin{figure}
    \centering
    \includegraphics[width=\columnwidth]{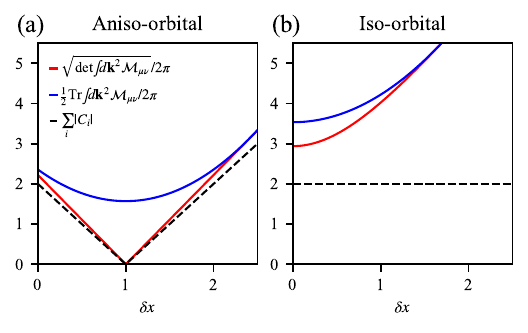}
    \caption{The trace and determinant of the integrated quantum metric for a) the aniso-orbital and b) the iso-orbital model. The minimum of the blue line corresponds to the minimal quantum metric. In the aniso-orbital case, the trace of integrated metric is finite, while the determinant vanishes, implying that superconductivity is unstable in the flat-band limit. For this particular model, it is unstable even without flattening due to the band dispersion being independent of $x$.}
    \label{fig:quantum_metric}
\end{figure}

So far we have only considered bounds for Chern dartboard insulators that satisfy the iso-orbital condition. As mentioned previously, in aniso-orbital models, the Chern number is only well-defined in specific bases (see Ref.~\cite{Chen2024} and the supplemental material~\cite{supplemental}). In any other basis, it is no longer an integer and depends linearly on the orbital displacement $\delta x$ along the high-symmetry line, $C'=C- \delta x$ (displacements perpendicular to the high-symmetry line have no effect on $C$). The bound in Eq.~\eqref{eq:final_bound} is derived in the basis which minimizes the trace of the integrated quantum metric. In the aniso-orbital case, since the Chern number is basis-dependent, it is essential that it is computed in this particular basis. Considering the Chern number can take any value, the safest choice, if the correct basis is unknown, is the trivial bound. This bound is the best possible in some cases (see example below), e.g. when the Berry curvature is identically zero in some basis. In other models, it is possible that $|C|>0$ in the basis minimizing the quantum metric (even though there exists another basis where $C=0$): this can occur when there are features in the Berry curvature that can not be flattened by basis changes. 

To illustrate the basis dependence, we consider an aniso-orbital model introduced in Ref.~\cite{Chen2024} with Hamiltonian $H^{\ssm an}(\vec{k}) = \vec{\sigma} \cdot\vec{d}^{an}(\vec{k})$, with
\begin{equation}
\begin{aligned}
\nonumber
    \vec{d}^{\ssm an}(\vec{k})= \Big[\cos(k_x)\sin(k_y),\, \sin(k_x)\sin (k_y),\, \cos(k_y)\Big] .
  \label{eq:H_aniso}
  \end{aligned}
\end{equation}

As can be seen from Fig.~\ref{fig:quantum_metric}.a, in this aniso-orbital model,
$\tfrac{1}{2\pi}\sqrt{{\rm det}\int d\mathbf{k}^2 \mathcal{M}_{\mu\nu}}$ and the Chern number vanish in the basis that minimizes ${\rm Tr}[\int d\vec{k}^2 \mathcal{M}_{\mu\nu}]$, even though the trace of the quantum metric remains nontrivial. In this case, focusing only on the trace of the quantum metric would incorrectly suggest a stable superconductor, highlighting that the determinant is the essential quantity in anisotropic systems.

In the iso-orbital case (Fig.~\ref{fig:quantum_metric}.b), $\tfrac{1}{2\pi}\sqrt{{\rm det}\int d\vec{k}^2 \mathcal{M}_{\mu\nu}}$ is non-zero when the trace of the quantum metric is minimized, and is bounded from below by the sum $\sum_i|C_i|$. This sum remains unchanged even if orbital positions are changed; computing the Chern number in any arbitrary basis gives the correct bound in iso-orbital models.

\section{Discussion}

We have shown that the geometric contribution to the superfluid weight can be bounded from below by the sum of the absolute values of quantized sub-BZ Chern numbers in isolated delicate bands. In the CDIs we consider, this results in a bound increasing linearly with the number of mirror planes. Even though the total Chern number vanishes, this can result in a high topological bound. Delicate topology requires a highly nontrivial Berry curvature, and thus quantum metric, which makes especially delicate flat bands, where the geometric contribution to the superfluid weight is dominant, particularly promising for superconductivity. In dispersive bands, the bound obtained from the sum of Chern numbers gives a lower estimate for the geometric contribution at low interactions, to be added to the conventional single-band result. 

We have emphasized the importance of the basis-independence of the Chern number when formulating bounds on the superfluid weight. In aniso-orbital models, the sub-BZ Chern number is basis-dependent, and can not be used reliably as a bound, since it is not necessarily a lower bound for the metric in an arbitrary basis. All aniso-orbital models are not equally trivial, however, and this calls for future work investigating potential basis-independent invariants that can differentiate between them. 

Our results apply to two-dimensional delicate bands, where we can define a Chern number in the irreducible BZ. In the models we consider, delicate topology is linked to mirror symmetry. In two dimensions, topological invariants in a sub-BZ can also be quantized in the presence of e.g. non-symmorphic mirror symmetries~\cite{Vaidya2025}. In higher dimensions, delicate topology can be protected by rotational symmetry, and is characterized by the returning Thouless pump or Hopf invariant~\cite{Nelson2022}. It remains an open question whether these other invariants can also be used to formulate bounds on the superfluid weight. 

Beyond the theoretical framework, it is natural to ask whether delicate topology can be realized experimentally in quantum materials. While no material candidates are yet known for realizing CDIs, several systems are suggested to host a nonzero returning Thouless pump (RTP), a hallmark of delicate topology. These include elemental $\alpha-\mathrm{Sn}$, half-Heusler alloys, pyrochlore iridates \cite{Zhu2024}, and strained hexagonal compounds such as $\mathrm{ZrTe}$, $\mathrm{WC}$, and $\mathrm{MoC}$ \cite{Nelson2022}.

Since quantum geometry is involved in various phenomena~\cite{Rossi2021,Liu2024,Yu2024,Torma2023} such as nonlinear transport including the nonlinear Hall effect~\cite{Jiang2025,Gao2014,Kaplan2024,Wang2023,Ulrich2025}, Bose-Einstein condensation in multiband models and flat bands~\cite{Torma2021,Lukin2023,Iskin2023,Lukin2023}, and relates in several ways to the conductivity and other transport  quantities~\cite{Resta2011,Resta2018,Antebi2024,Komissarov2025,Verma2024,Kruchkov2023,Mera2022,Mitscherling2022,Bouzerar2022},
delicate bands might be promising for the observation of correlated phenomena beyond superconductivity.

\begin{acknowledgments}
K.-E.H. is grateful for support by an ETH Zürich Postdoctoral Fellowship.
\end{acknowledgments}

\section*{Data availability}

The data that support the findings of this article are openly
available~\cite{data}.

\pagebreak
\widetext
\setcounter{equation}{0}
\setcounter{figure}{0}
\setcounter{table}{0}
\setcounter{section}{0}
\setcounter{page}{1}
\makeatletter
\renewcommand{\theequation}{S\arabic{equation}}
\renewcommand{\thefigure}{S\arabic{figure}}
\renewcommand{\bibnumfmt}[1]{[S#1]}
\renewcommand{\citenumfont}[1]{S#1}

\begin{center}
  {\bf Supplemental material: Superfluid stiffness of superconductors
    with delicate topology} \\
  \vspace{0.5cm}
  Tijan Prijon, Sebastian D. Huber and Kukka-Emilia Huhtinen \\
  {\it 
    Institute for Theoretical Physics, ETH Zurich, 8093 Zürich, Switzerland
  }
\end{center}

\section{Basis (in)dependence of the Chern number} \label{app:basis-dependence}

The basis independence of the Chern number over an irreducible BZ fully bounded by a line at which the $n_b$ bands of interest occupy the same $n_b$ orbitals is easy to check. This includes the models with more than one mirror considered in the main text.
With only one mirror, however, the boundary of the half Brillouin contains two HSLs at which the mirror representations, and thus the occupied orbitals, can be different. Whether they are the same (iso-orbital) or different (aniso-orbital) has a drastic impact on the basis-independence and quantization of the Chern number. 

For concreteness, let us focus on a Hamiltonian $H_{\vec{k}}$ with a
mirror symmetry 
\begin{equation}
  M_y H_{(k_x,k_y)} M_{y}^{-1} = H_{(k_x,-k_y)},
\end{equation}
where $M_y$ is picked to be diagonal. The two HSLs dividing
the BZ into halves are $k_y=0$ and $k_y=\pi$. We now focus
on the integral of the Berry curvature over the irreducible BZ when a
basis transformation $\mathcal{U}_{\vec{k}}\to
\widetilde{\mathcal{U}}_{\vec{k}} = V_{\vec{k}}\mathcal{U}_{\vec{k}}$,
$[V_{\vec{k}}]_{\alpha\beta} = e^{i\vec{k}\cdot
  \vec{r}^{\alpha}}\delta_{\alpha\beta}$ is performed. The basis
transformation shifts the position of orbital $\alpha$ by
$\vec{r}^{\alpha}$ while the tight-binding parameters are kept
unchanged. In other words, it is a change of Fourier transformation
convention, which includes effective redefinitions of the unit cell.

The Berry connection $\mathcal{A}_{\vec{k}} = i{\rm Tr}[\mathcal{U}_{\vec{k}}^{\dag}\nabla \mathcal{U}_{\vec{k}}]^{\vphantom{\dag}}$ transforms as
\begin{equation}
  \begin{aligned}
    \widetilde{\mathcal{A}}_{\vec{k}} =& i{\rm
      Tr}[\widetilde{\mathcal{U}}_{\vec{k}}^{\dag} \nabla
      \widetilde{\mathcal{U}}_{\vec{k}}^{\vphantom{\dag}}] = i{\rm
      Tr}[\mathcal{U}_{\vec{k}}^{\dag} V_{\vec{k}}^{\dag} \nabla 
    (V_{\vec{k}}^{\vphantom{\dag}}
    \mathcal{U}_{\vec{k}}^{\vphantom{\dag}})] 
    = \mathcal{A}_{\vec{k}} -{\rm Tr}[\vec{R} \mathcal{U_{\vec{k}}}^{\vphantom{\dag}}
    \mathcal{U}_{\vec{k}}^{\dag}], 
  \end{aligned}
\end{equation}
where $[R_{\mu}]_{\alpha\beta} =
r^{\alpha}_{\mu}\delta_{\alpha\beta}$. The change in the integral of
the Berry curvature is then
\begin{equation}
  \begin{aligned}
    &\Delta C = \widetilde{C} - C = -\frac{1}{2\pi} \int_{\partial{\rm
        irBz}} d\vec{k}\cdot {\rm
      Tr}[\vec{R}\mathcal{U}_{\vec{k}}^{\vphantom{\dag}}\mathcal{U}_{\vec{k}}^{\dag}]
    \\
    =& \frac{1}{2\pi} \int_{0}^{\pi}dk_y \left({\rm Tr}[R_y
      \mathcal{U}_{(-\pi,k_y)}^{\vphantom{\dag}}
      \mathcal{U}_{(-\pi,k_y)}^{\dag}] - {\rm Tr}[R_y
      \mathcal{U}_{(\pi,k_y)}^{\vphantom{\dag}}
      \mathcal{U}_{(\pi,k_y)}^{\dag}]\right) \\
    +& \frac{1}{2\pi} \int_{-\pi}^{\pi}dk_x \left({\rm Tr}[R_x
      \mathcal{U}_{(k_x,\pi)}^{\vphantom{\dag}}
      \mathcal{U}_{(k_x,\pi)}^{\dag}] - {\rm Tr}[R_x
      \mathcal{U}_{(k_x,0)}^{\vphantom{\dag}}
      \mathcal{U}_{(k_x,0)}^{\dag}]\right) \label{eq.basis_transform}
  \end{aligned} 
\end{equation}
Regardless of basis,
$[\mathcal{U}_{(-\pi,k_y)}^{\vphantom{\dag}}\mathcal{U}_{(-\pi,k_y)}^{\dag}]_{\alpha,\alpha}
= \sum_{n\in\mathcal{B}} |\braket{\alpha}{n_{(-\pi,k_y)}}|^2 =
\sum_{n\in\mathcal{B}} |\braket{\alpha}{n_{(\pi,k_y)}}|^2 =
    [\mathcal{U}_{(\pi,k_y)}^{\vphantom{\dag}}\mathcal{U}_{(\pi,k_y)}^{\dag}]_{\alpha,\alpha}$
    since basis transformations do not affect the absolute values of
    Bloch function components, which are therefore always
    periodic. As a consequence, the second line of Eq.~\eqref{eq.basis_transform} vanishes.

On the third line of Eq.~\eqref{eq.basis_transform}, we can use the
fact that the $n_b$ bands in 
$\mathcal{B}$ occupy exactly $n_b$ orbitals at the HSLs, and thus ${\rm
  Tr}[R_x\mathcal{U}_{(k_x,k_y)}^{\vphantom{\dag}}
  \mathcal{U}_{(k_x,k_y)}^{\dag} ] = {\rm
  Tr}[R_{x,k_y}^{\mathcal{B}}\mathcal{V}_{(k_x,k_y)}^{\vphantom{\dag}}
  \mathcal{V}_{(k_x,k_y)}^{\dag} ] = {\rm Tr}[R_{x,k_y}^{\mathcal{B}}]$. Here,
$R_{x,k_y}^{\mathcal{B}}$ contains only those orbitals which are
occupied by the bands in $\mathcal{B}$ along the line specified by the value of $k_y$. The change in
Chern number is thus
\begin{equation}
  \Delta C = {\rm Tr}[R_{x,\pi}^{\mathcal{B}}] - {\rm Tr}[R_{x,0}^{\mathcal{B}}].
\end{equation}
If the iso-orbitality condition is fulfilled, $R_{x,\pi}^{\mathcal{B}}
= R_{x,0}^{\mathcal{B}}$ since the same orbitals are occupied at $k_y=0$ and $k_y=\pi$, and the Chern number is indeed always well-defined
and independent of basis. In aniso-orbital models, however, different
orbitals are occupied, and the Chern number is only a well-defined integer in particular bases, otherwise depending linearly on
relative shifts in orbital positions along $x$. In periodic bases, where it is always quantized, its value will depend on the choice of unit cell. In this sense, the sub-BZ Chern number behaves
similarly to a 1D winding number in aniso-orbital systems, whereas it is
a well-defined basis-independent Chern number in iso-orbital systems. 

\section{Superfluid weight in system with uniform pairing} \label{app:sfw_upc}

As mentioned in the main text, the superfluid weight in the basis where $\delta\Delta_{\alpha}/\delta A_{\mu}=0$ is given by
\begin{equation}
  \begin{aligned}
    D_{\mu\nu} =& \frac{1}{N}\sum_{\vec{k},i,j} 
    \frac{n_F(E_j)-n_F(E_i)}{E_i-E_j}\\
    &\bigg(
    \bra{\psi_i}\partial_{\mu}H_{\rm \scriptscriptstyle  BdG}(\vec{k}) \ket{\psi_j}
    \bra{\psi_j}\partial_{\nu}H_{\rm \scriptscriptstyle  BdG}(\vec{k})\ket{\psi_i}  
    - \bra{\psi_i}\partial_{\mu} H_{\rm \scriptscriptstyle  BdG}(\vec{k})\gamma^z
    \ket{\psi_j} \bra{\psi_j} \partial_{\nu}H_{\rm \scriptscriptstyle  BdG}(\vec{k})\gamma^z
    \ket{\psi_i} \bigg), \label{eq.ds}
  \end{aligned}
\end{equation}
where $E_i$ and $\ket{\psi_i}$ are the eigenvalues and eigenvectors of
$H_{\rm \scriptscriptstyle  BdG}$, respectively, and $\gamma^z = \sigma_z\otimes
\mathbb{1}_M$, with $\mathbb{1}_M$ the $M\times M$ identity matrix. We
use the shorthand $\partial_{\mu}\equiv \partial/\partial_{k_{\mu}}$. In
the prefactor, $n_F(E) = 1/(e^{\beta E}+1)$ is the Fermi-Dirac
distribution. When $E=E_i=E_j$, the prefactor should is
$-\partial_En_F(E)$.

The determination of $E_i$ and $\ket{\psi_i}$ often requires numerical
diagonalization of $H_{BdG}$. However, if we assume the uniform
pairing condition (UPC) $\Delta_{\alpha}=\Delta$ holds, the
eigenvalues come in pairs $E_n^{\pm}=\pm E_n=\pm
\sqrt{(\varepsilon_n(\vec{k})-\mu)^2+|\Delta|^2}$ and $\ket{\psi_n^+}
= (u_n^+\ket{+}+u_n^-\ket{-})\otimes\ket{n_{\vec{k}}}$, $\ket{\psi_n^-} =
(-u_n^-\ket{+} + u_n^+\ket{-})\otimes\ket{n_{\vec{k}}}$, where
\begin{equation}
  u_n^{\pm}=\frac{1}{\sqrt{2}}\sqrt{1 \pm \frac{\varepsilon_n-\mu}{E_n}},
\end{equation}
and $\ket{n_{\vec{k}}}$ is the eigenvector of $H_{\vec{k}}$
corresponding to the eigenvalue $\varepsilon_n(\vec{k})$. These
expressions can be plugged into Eq.~\eqref{eq.ds} to obtain $D=D^{\rm
  conv} + D^{\rm geom}$,
\begin{equation}
  \begin{aligned}
    D^{\rm conv}_{\mu\nu} =& \frac{|\Delta|^2}{N}\sum_{\vec{k}n}
    \frac{1}{E_n^2}\left[ \frac{\tanh(\beta
        E_n/2)}{E_n}-\frac{\beta}{2\cosh^2(\beta E_n/2)} \right]
    \partial_{\mu}\varepsilon_n\partial_{\nu}\varepsilon_n ,\\ 
    D^{\rm geom}_{\mu\nu} =& \frac{|\Delta|^2}{N}\sum_{\vec{k},m\neq n}\left[\frac{\tanh(\beta
      E_m/2)}{E_m}-\frac{\tanh(\beta E_n/2)}{E_n}\right]
    \frac{\varepsilon_n-\varepsilon_m}{\varepsilon_n+\varepsilon_m-2\mu}
    (\braket{\partial_{\mu}m}{n} \braket{n}{\partial_{\nu}m}+{\rm
      H.c.}). \label{eq.upc_ds}
  \end{aligned}
\end{equation}
We have used here that
$\bra{n_{\vec{k}}}\partial_{\mu}H_{\vec{k}}\ket{m_{\vec{k}}} =
\partial_{\mu}\varepsilon_n(\vec{k})\delta_{nm} -
(\varepsilon_m(\vec{k})-\varepsilon_n(\vec{k}))\braket{\partial_{\mu}n_{\vec{k}}}{m_{\vec{k}}}$. 
Note that there are no additional terms because of the basis we work in. If the basis is not
fixed, additional terms need to be taken into account.

Under the isolated band assumption $\varepsilon_n=\varepsilon_{\overline{n}}$, 
$|\varepsilon_n-\varepsilon_m|\gg |\varepsilon_n-\mu|$ and
$|\varepsilon_n-\varepsilon_m|\gg|\Delta|$ 
 for all $n\in\mathcal{B}$  
and $m\notin\mathcal{B}$, 
\begin{equation}
  D^{\rm geom}_{\mu\nu} = \frac{(2\pi|\Delta|)^2}{N}\int_{\rm B.z.} d\vec{k}^2
  f^{\mathcal{B}}(\vec{k}) \mathcal{M}^{\mathcal{B}}(\vec{k}),
\end{equation}
where $f^{\mathcal{B}}(\vec{k}) = 2\tanh(\beta
E_{\overline{n}}(\vec{k})/2)/E_{\overline{n}}(\vec{k})/(2\pi)^2$. In the thermodynamic limit, we replace the sum $(1/N)\sum_{\vec{k}}$ by an integral $(1/2\pi)^2 \int_{\rm B.z} d\vec{k}^2$ over the first BZ, and obtain Eq.~(7) in the main text.

\section{Models}
\label{app:models}

The Hamiltonians for the iso-orbital Chern dartboard models~\cite{Chen2024} studied in the main text are
\begin{align}
&\begin{aligned}
    \vec{d}^1(\vec{k})
      = \Bigl( &\tfrac{1}{2}\bigl(1+\cos(k_x)\bigr)\sin(2k_y),\, \\ &\sin(k_x) \sin(k_y),\, \\
  &m+\tfrac{1}{2}\bigl(1+\cos(k_x)\bigr)\bigl(\cos(2k_y)-1\bigr) \Bigr), 
  \label{eq:H1}
  \end{aligned}\\
&\begin{aligned}
     \vec{d}^2(\vec{k})
      = \Bigl(& -\sin (k_x)\sin(2k_y),\, \\ &\sin(2k_x)\sin(k_y),  \\
  & m+\cos(2k_x)+\cos(2k_y) \Bigr), 
  \label{eq:H2}
  \end{aligned} \\
&\begin{aligned}
\vec{d}^3(\vec{k})
= \Bigl(&\sin(\tfrac{5}{2}k_x)\sin(\tfrac{\sqrt{3}}{2}k_y)
       - \sin(2k_x)\sin(\sqrt{3}k_y)
        + \sin(\tfrac{1}{2}k_x)\sin(\tfrac{3\sqrt{3}}{2}k_y), \\
        &-\cos(\tfrac{5}{2}k_x) \sin(\tfrac{\sqrt{3}}{2}k_y)- \cos(2k_x)\sin(\sqrt{3}k_y) + \cos(\tfrac{1}{2}k_x)\sin(\tfrac{3\sqrt{3}}{2}k_y), \\
&\tfrac{4}{9}m+\sum_{a=1}^{6}(e^{i\vec{t}(a)\cdot\vec{k}}
       + \tfrac{1}{2}e^{2i\vec{t}(a)\cdot\vec{k}})\Bigr),
\end{aligned}
\label{eq:H3} \\
&
\begin{aligned}
\vec{d}^4(\vec{k})
&= \Bigl( -\sin k_x \sin(4k_y)+\sin(4k_x)\sin(k_y), \\
&\quad \sin(2k_x)\sin(4k_y)-\sin(4k_x)\sin(2k_y), \\
&\quad \tfrac{1}{2}m+\cos(2k_x)+\cos(2k_y)+\cos(4k_x)+\cos(4k_y)+4 \cos(k_x) \cos(k_y)-1 \Bigr). 
\end{aligned}
\label{eq:H4}
\end{align}
In $\vec{d}^3$, $\vec{t}(a)=\sqrt{3}\,\Big[\cos(\frac{\pi a}{3}-\frac{\pi}{6}),\;
\sin(\frac{\pi a}{3}-\frac{\pi}{6})\Big].$


\begin{thebibliography}{68}%
\makeatletter
\providecommand \@ifxundefined [1]{%
 \@ifx{#1\undefined}
}%
\providecommand \@ifnum [1]{%
 \ifnum #1\expandafter \@firstoftwo
 \else \expandafter \@secondoftwo
 \fi
}%
\providecommand \@ifx [1]{%
 \ifx #1\expandafter \@firstoftwo
 \else \expandafter \@secondoftwo
 \fi
}%
\providecommand \natexlab [1]{#1}%
\providecommand \enquote  [1]{``#1''}%
\providecommand \bibnamefont  [1]{#1}%
\providecommand \bibfnamefont [1]{#1}%
\providecommand \citenamefont [1]{#1}%
\providecommand \href@noop [0]{\@secondoftwo}%
\providecommand \href [0]{\begingroup \@sanitize@url \@href}%
\providecommand \@href[1]{\@@startlink{#1}\@@href}%
\providecommand \@@href[1]{\endgroup#1\@@endlink}%
\providecommand \@sanitize@url [0]{\catcode `\\12\catcode `\$12\catcode
  `\&12\catcode `\#12\catcode `\^12\catcode `\_12\catcode `\%12\relax}%
\providecommand \@@startlink[1]{}%
\providecommand \@@endlink[0]{}%
\providecommand \url  [0]{\begingroup\@sanitize@url \@url }%
\providecommand \@url [1]{\endgroup\@href {#1}{\urlprefix }}%
\providecommand \urlprefix  [0]{URL }%
\providecommand \Eprint [0]{\href }%
\providecommand \doibase [0]{https://doi.org/}%
\providecommand \selectlanguage [0]{\@gobble}%
\providecommand \bibinfo  [0]{\@secondoftwo}%
\providecommand \bibfield  [0]{\@secondoftwo}%
\providecommand \translation [1]{[#1]}%
\providecommand \BibitemOpen [0]{}%
\providecommand \bibitemStop [0]{}%
\providecommand \bibitemNoStop [0]{.\EOS\space}%
\providecommand \EOS [0]{\spacefactor3000\relax}%
\providecommand \BibitemShut  [1]{\csname bibitem#1\endcsname}%
\let\auto@bib@innerbib\@empty
\bibitem [{\citenamefont {Mielke}\ \emph {et~al.}(2021)\citenamefont {Mielke},
  \citenamefont {Qin}, \citenamefont {Yin}, \citenamefont {Nakamura},
  \citenamefont {Das}, \citenamefont {Guo}, \citenamefont {Khasanov},
  \citenamefont {Chang}, \citenamefont {Wang}, \citenamefont {Jia},
  \citenamefont {Nakatsuji}, \citenamefont {Amato}, \citenamefont {Luetkens},
  \citenamefont {Xu}, \citenamefont {Hasan},\ and\ \citenamefont
  {Guguchia}}]{Mielke2021}%
  \BibitemOpen
  \bibfield  {author} {\bibinfo {author} {\bibfnamefont {C.}~\bibnamefont
  {Mielke}}, \bibinfo {author} {\bibfnamefont {Y.}~\bibnamefont {Qin}},
  \bibinfo {author} {\bibfnamefont {J.-X.}\ \bibnamefont {Yin}}, \bibinfo
  {author} {\bibfnamefont {H.}~\bibnamefont {Nakamura}}, \bibinfo {author}
  {\bibfnamefont {D.}~\bibnamefont {Das}}, \bibinfo {author} {\bibfnamefont
  {K.}~\bibnamefont {Guo}}, \bibinfo {author} {\bibfnamefont {R.}~\bibnamefont
  {Khasanov}}, \bibinfo {author} {\bibfnamefont {J.}~\bibnamefont {Chang}},
  \bibinfo {author} {\bibfnamefont {Z.~Q.}\ \bibnamefont {Wang}}, \bibinfo
  {author} {\bibfnamefont {S.}~\bibnamefont {Jia}}, \bibinfo {author}
  {\bibfnamefont {S.}~\bibnamefont {Nakatsuji}}, \bibinfo {author}
  {\bibfnamefont {A.}~\bibnamefont {Amato}}, \bibinfo {author} {\bibfnamefont
  {H.}~\bibnamefont {Luetkens}}, \bibinfo {author} {\bibfnamefont
  {G.}~\bibnamefont {Xu}}, \bibinfo {author} {\bibfnamefont {M.~Z.}\
  \bibnamefont {Hasan}},\ and\ \bibinfo {author} {\bibfnamefont
  {Z.}~\bibnamefont {Guguchia}},\ }\bibfield  {title} {\bibinfo {title}
  {Nodeless kagome superconductivity in
  {${\mathrm{LaRu}}_{3}{\mathrm{Si}}_{2}$}},\ }\href
  {https://doi.org/10.1103/PhysRevMaterials.5.034803} {\bibfield  {journal}
  {\bibinfo  {journal} {Phys. Rev. Mater.}\ }\textbf {\bibinfo {volume} {5}},\
  \bibinfo {pages} {034803} (\bibinfo {year} {2021})}\BibitemShut {NoStop}%
\bibitem [{\citenamefont {Khasanov}\ \emph {et~al.}(2024)\citenamefont
  {Khasanov}, \citenamefont {Ruan}, \citenamefont {Shi}, \citenamefont {Chen},
  \citenamefont {Luetkens}, \citenamefont {Ren},\ and\ \citenamefont
  {Guguchia}}]{Khasanov2024}%
  \BibitemOpen
  \bibfield  {author} {\bibinfo {author} {\bibfnamefont {R.}~\bibnamefont
  {Khasanov}}, \bibinfo {author} {\bibfnamefont {B.-B.}\ \bibnamefont {Ruan}},
  \bibinfo {author} {\bibfnamefont {Y.-Q.}\ \bibnamefont {Shi}}, \bibinfo
  {author} {\bibfnamefont {G.-F.}\ \bibnamefont {Chen}}, \bibinfo {author}
  {\bibfnamefont {H.}~\bibnamefont {Luetkens}}, \bibinfo {author}
  {\bibfnamefont {Z.-A.}\ \bibnamefont {Ren}},\ and\ \bibinfo {author}
  {\bibfnamefont {Z.}~\bibnamefont {Guguchia}},\ }\bibfield  {title} {\bibinfo
  {title} {Tuning of the flat band and its impact on superconductivity in
  {Mo5SI3-xPx}},\ }\href {https://doi.org/10.1038/s41467-024-46514-2}
  {\bibfield  {journal} {\bibinfo  {journal} {Nature Communications}\ }\textbf
  {\bibinfo {volume} {15}},\ \bibinfo {pages} {2197} (\bibinfo {year}
  {2024})}\BibitemShut {NoStop}%
\bibitem [{\citenamefont {Cao}\ \emph {et~al.}(2018)\citenamefont {Cao},
  \citenamefont {Fatemi}, \citenamefont {Demir}, \citenamefont {Fang},
  \citenamefont {Tomarken}, \citenamefont {Luo}, \citenamefont
  {Sanchez-Yamagishi}, \citenamefont {Watanabe}, \citenamefont {Taniguchi},
  \citenamefont {Kaxiras}, \citenamefont {Ashoori},\ and\ \citenamefont
  {Jarillo-Herrero}}]{Cao2018}%
  \BibitemOpen
  \bibfield  {author} {\bibinfo {author} {\bibfnamefont {Y.}~\bibnamefont
  {Cao}}, \bibinfo {author} {\bibfnamefont {V.}~\bibnamefont {Fatemi}},
  \bibinfo {author} {\bibfnamefont {A.}~\bibnamefont {Demir}}, \bibinfo
  {author} {\bibfnamefont {S.}~\bibnamefont {Fang}}, \bibinfo {author}
  {\bibfnamefont {S.~L.}\ \bibnamefont {Tomarken}}, \bibinfo {author}
  {\bibfnamefont {J.~Y.}\ \bibnamefont {Luo}}, \bibinfo {author} {\bibfnamefont
  {J.~D.}\ \bibnamefont {Sanchez-Yamagishi}}, \bibinfo {author} {\bibfnamefont
  {K.}~\bibnamefont {Watanabe}}, \bibinfo {author} {\bibfnamefont
  {T.}~\bibnamefont {Taniguchi}}, \bibinfo {author} {\bibfnamefont
  {E.}~\bibnamefont {Kaxiras}}, \bibinfo {author} {\bibfnamefont {R.~C.}\
  \bibnamefont {Ashoori}},\ and\ \bibinfo {author} {\bibfnamefont
  {P.}~\bibnamefont {Jarillo-Herrero}},\ }\bibfield  {title} {\bibinfo {title}
  {Correlated insulator behaviour at half-filling in magic-angle graphene
  superlattices},\ }\href {https://doi.org/10.1038/nature26154} {\bibfield
  {journal} {\bibinfo  {journal} {Nature}\ }\textbf {\bibinfo {volume} {556}},\
  \bibinfo {pages} {80} (\bibinfo {year} {2018})}\BibitemShut {NoStop}%
\bibitem [{\citenamefont {Lu}\ \emph {et~al.}(2019)\citenamefont {Lu},
  \citenamefont {Stepanov}, \citenamefont {Yang}, \citenamefont {Xie},
  \citenamefont {Aamir}, \citenamefont {Das}, \citenamefont {Urgell},
  \citenamefont {Watanabe}, \citenamefont {Taniguchi}, \citenamefont {Zhang},
  \citenamefont {Bachtold}, \citenamefont {MacDonald},\ and\ \citenamefont
  {Efetov}}]{Lu2019}%
  \BibitemOpen
  \bibfield  {author} {\bibinfo {author} {\bibfnamefont {X.}~\bibnamefont
  {Lu}}, \bibinfo {author} {\bibfnamefont {P.}~\bibnamefont {Stepanov}},
  \bibinfo {author} {\bibfnamefont {W.}~\bibnamefont {Yang}}, \bibinfo {author}
  {\bibfnamefont {M.}~\bibnamefont {Xie}}, \bibinfo {author} {\bibfnamefont
  {M.~A.}\ \bibnamefont {Aamir}}, \bibinfo {author} {\bibfnamefont
  {I.}~\bibnamefont {Das}}, \bibinfo {author} {\bibfnamefont {C.}~\bibnamefont
  {Urgell}}, \bibinfo {author} {\bibfnamefont {K.}~\bibnamefont {Watanabe}},
  \bibinfo {author} {\bibfnamefont {T.}~\bibnamefont {Taniguchi}}, \bibinfo
  {author} {\bibfnamefont {G.}~\bibnamefont {Zhang}}, \bibinfo {author}
  {\bibfnamefont {A.}~\bibnamefont {Bachtold}}, \bibinfo {author}
  {\bibfnamefont {A.~H.}\ \bibnamefont {MacDonald}},\ and\ \bibinfo {author}
  {\bibfnamefont {D.~K.}\ \bibnamefont {Efetov}},\ }\bibfield  {title}
  {\bibinfo {title} {Superconductors, orbital magnets and correlated states in
  magic-angle bilayer graphene},\ }\href
  {https://doi.org/10.1038/s41586-019-1695-0} {\bibfield  {journal} {\bibinfo
  {journal} {Nature}\ }\textbf {\bibinfo {volume} {574}},\ \bibinfo {pages}
  {653} (\bibinfo {year} {2019})}\BibitemShut {NoStop}%
\bibitem [{\citenamefont {Hao}\ \emph {et~al.}(2021)\citenamefont {Hao},
  \citenamefont {Zimmerman}, \citenamefont {Ledwith}, \citenamefont {Khalaf},
  \citenamefont {Najafabadi}, \citenamefont {Watanabe}, \citenamefont
  {Taniguchi}, \citenamefont {Vishwanath},\ and\ \citenamefont
  {Kim}}]{Hao2021}%
  \BibitemOpen
  \bibfield  {author} {\bibinfo {author} {\bibfnamefont {Z.}~\bibnamefont
  {Hao}}, \bibinfo {author} {\bibfnamefont {A.~M.}\ \bibnamefont {Zimmerman}},
  \bibinfo {author} {\bibfnamefont {P.}~\bibnamefont {Ledwith}}, \bibinfo
  {author} {\bibfnamefont {E.}~\bibnamefont {Khalaf}}, \bibinfo {author}
  {\bibfnamefont {D.~H.}\ \bibnamefont {Najafabadi}}, \bibinfo {author}
  {\bibfnamefont {K.}~\bibnamefont {Watanabe}}, \bibinfo {author}
  {\bibfnamefont {T.}~\bibnamefont {Taniguchi}}, \bibinfo {author}
  {\bibfnamefont {A.}~\bibnamefont {Vishwanath}},\ and\ \bibinfo {author}
  {\bibfnamefont {P.}~\bibnamefont {Kim}},\ }\bibfield  {title} {\bibinfo
  {title} {Electric field–tunable superconductivity in alternating-twist
  magic-angle trilayer graphene},\ }\href
  {https://doi.org/10.1126/science.abg0399} {\bibfield  {journal} {\bibinfo
  {journal} {Science}\ }\textbf {\bibinfo {volume} {371}},\ \bibinfo {pages}
  {1133} (\bibinfo {year} {2021})},\ \Eprint
  {https://arxiv.org/abs/https://www.science.org/doi/pdf/10.1126/science.abg0399}
  {https://www.science.org/doi/pdf/10.1126/science.abg0399} \BibitemShut
  {NoStop}%
\bibitem [{\citenamefont {Park}\ \emph {et~al.}(2021)\citenamefont {Park},
  \citenamefont {Cao}, \citenamefont {Watanabe}, \citenamefont {Taniguchi},\
  and\ \citenamefont {Jarillo-Herrero}}]{Park2021}%
  \BibitemOpen
  \bibfield  {author} {\bibinfo {author} {\bibfnamefont {J.~M.}\ \bibnamefont
  {Park}}, \bibinfo {author} {\bibfnamefont {Y.}~\bibnamefont {Cao}}, \bibinfo
  {author} {\bibfnamefont {K.}~\bibnamefont {Watanabe}}, \bibinfo {author}
  {\bibfnamefont {T.}~\bibnamefont {Taniguchi}},\ and\ \bibinfo {author}
  {\bibfnamefont {P.}~\bibnamefont {Jarillo-Herrero}},\ }\bibfield  {title}
  {\bibinfo {title} {Tunable strongly coupled superconductivity in magic-angle
  twisted trilayer graphene},\ }\href
  {https://doi.org/10.1038/s41586-021-03192-0} {\bibfield  {journal} {\bibinfo
  {journal} {Nature}\ }\textbf {\bibinfo {volume} {590}},\ \bibinfo {pages}
  {249} (\bibinfo {year} {2021})}\BibitemShut {NoStop}%
\bibitem [{\citenamefont {Park}\ \emph {et~al.}(2022)\citenamefont {Park},
  \citenamefont {Cao}, \citenamefont {Xia}, \citenamefont {Sun}, \citenamefont
  {Watanabe}, \citenamefont {Taniguchi},\ and\ \citenamefont
  {Jarillo-Herrero}}]{Park2022}%
  \BibitemOpen
  \bibfield  {author} {\bibinfo {author} {\bibfnamefont {J.~M.}\ \bibnamefont
  {Park}}, \bibinfo {author} {\bibfnamefont {Y.}~\bibnamefont {Cao}}, \bibinfo
  {author} {\bibfnamefont {L.-Q.}\ \bibnamefont {Xia}}, \bibinfo {author}
  {\bibfnamefont {S.}~\bibnamefont {Sun}}, \bibinfo {author} {\bibfnamefont
  {K.}~\bibnamefont {Watanabe}}, \bibinfo {author} {\bibfnamefont
  {T.}~\bibnamefont {Taniguchi}},\ and\ \bibinfo {author} {\bibfnamefont
  {P.}~\bibnamefont {Jarillo-Herrero}},\ }\bibfield  {title} {\bibinfo {title}
  {Robust superconductivity in magic-angle multilayer graphene family},\ }\href
  {https://doi.org/10.1038/s41563-022-01287-1} {\bibfield  {journal} {\bibinfo
  {journal} {Nature Materials}\ }\textbf {\bibinfo {volume} {21}},\ \bibinfo
  {pages} {877} (\bibinfo {year} {2022})}\BibitemShut {NoStop}%
\bibitem [{\citenamefont {Zhou}\ \emph {et~al.}(2021)\citenamefont {Zhou},
  \citenamefont {Xie}, \citenamefont {Taniguchi}, \citenamefont {Watanabe},\
  and\ \citenamefont {Young}}]{Zhou2021b}%
  \BibitemOpen
  \bibfield  {author} {\bibinfo {author} {\bibfnamefont {H.}~\bibnamefont
  {Zhou}}, \bibinfo {author} {\bibfnamefont {T.}~\bibnamefont {Xie}}, \bibinfo
  {author} {\bibfnamefont {T.}~\bibnamefont {Taniguchi}}, \bibinfo {author}
  {\bibfnamefont {K.}~\bibnamefont {Watanabe}},\ and\ \bibinfo {author}
  {\bibfnamefont {A.~F.}\ \bibnamefont {Young}},\ }\bibfield  {title} {\bibinfo
  {title} {Superconductivity in rhombohedral trilayer graphene},\ }\href
  {https://doi.org/10.1038/s41586-021-03926-0} {\bibfield  {journal} {\bibinfo
  {journal} {Nature}\ }\textbf {\bibinfo {volume} {598}},\ \bibinfo {pages}
  {434} (\bibinfo {year} {2021})}\BibitemShut {NoStop}%
\bibitem [{\citenamefont {Schrieffer}(1964)}]{Schrieffer1964}%
  \BibitemOpen
  \bibfield  {author} {\bibinfo {author} {\bibfnamefont {J.~R.}\ \bibnamefont
  {Schrieffer}},\ }\href@noop {} {\emph {\bibinfo {title} {Theory of
  Superconductivity}}}\ (\bibinfo  {publisher} {Addison-Wesley},\ \bibinfo
  {address} {Reading, MA},\ \bibinfo {year} {1964})\BibitemShut {NoStop}%
\bibitem [{\citenamefont {Scalapino}\ \emph {et~al.}(1993)\citenamefont
  {Scalapino}, \citenamefont {White},\ and\ \citenamefont
  {Zhang}}]{Scalapino1993}%
  \BibitemOpen
  \bibfield  {author} {\bibinfo {author} {\bibfnamefont {D.~J.}\ \bibnamefont
  {Scalapino}}, \bibinfo {author} {\bibfnamefont {S.~R.}\ \bibnamefont
  {White}},\ and\ \bibinfo {author} {\bibfnamefont {S.}~\bibnamefont {Zhang}},\
  }\bibfield  {title} {\bibinfo {title} {Insulator, metal, or superconductor:
  The criteria},\ }\href {https://doi.org/10.1103/PhysRevB.47.7995} {\bibfield
  {journal} {\bibinfo  {journal} {Phys. Rev. B}\ }\textbf {\bibinfo {volume}
  {47}},\ \bibinfo {pages} {7995} (\bibinfo {year} {1993})}\BibitemShut
  {NoStop}%
\bibitem [{\citenamefont {Berezinskii}(1971)}]{Berezinskii1970}%
  \BibitemOpen
  \bibfield  {author} {\bibinfo {author} {\bibfnamefont {V.~L.}\ \bibnamefont
  {Berezinskii}},\ }\bibfield  {title} {\bibinfo {title} {{Destruction of
  Long-range Order in One-dimensional and Two-dimensional Systems having a
  Continuous Symmetry Group I. Classical Systems}},\ }\href@noop {} {\bibfield
  {journal} {\bibinfo  {journal} {Sov. Phys. JETP}\ }\textbf {\bibinfo {volume}
  {32}},\ \bibinfo {pages} {493} (\bibinfo {year} {1971})}\BibitemShut
  {NoStop}%
\bibitem [{\citenamefont {Kosterlitz}\ and\ \citenamefont
  {Thouless}(1972)}]{Kosterlitz72}%
  \BibitemOpen
  \bibfield  {author} {\bibinfo {author} {\bibfnamefont {J.~M.}\ \bibnamefont
  {Kosterlitz}}\ and\ \bibinfo {author} {\bibfnamefont {D.~J.}\ \bibnamefont
  {Thouless}},\ }\bibfield  {title} {\bibinfo {title} {Long range order and
  metastability in two dimensional solids and superfluids. (application of
  dislocation theory)},\ }\href {https://doi.org/10.1088/0022-3719/5/11/002}
  {\bibfield  {journal} {\bibinfo  {journal} {Journal of Physics C: Solid State
  Physics}\ }\textbf {\bibinfo {volume} {5}},\ \bibinfo {pages} {L124}
  (\bibinfo {year} {1972})}\BibitemShut {NoStop}%
\bibitem [{\citenamefont {Kosterlitz}\ and\ \citenamefont
  {Thouless}(1973)}]{Kosterlitz73}%
  \BibitemOpen
  \bibfield  {author} {\bibinfo {author} {\bibfnamefont {J.~M.}\ \bibnamefont
  {Kosterlitz}}\ and\ \bibinfo {author} {\bibfnamefont {D.~J.}\ \bibnamefont
  {Thouless}},\ }\bibfield  {title} {\bibinfo {title} {Ordering, metastability
  and phase transitions in two-dimensional systems},\ }\href
  {https://doi.org/10.1088/0022-3719/6/7/010} {\bibfield  {journal} {\bibinfo
  {journal} {Journal of Physics C: Solid State Physics}\ }\textbf {\bibinfo
  {volume} {6}},\ \bibinfo {pages} {1181} (\bibinfo {year} {1973})}\BibitemShut
  {NoStop}%
\bibitem [{\citenamefont {Nelson}\ and\ \citenamefont
  {Kosterlitz}(1977)}]{Nelson77}%
  \BibitemOpen
  \bibfield  {author} {\bibinfo {author} {\bibfnamefont {D.~R.}\ \bibnamefont
  {Nelson}}\ and\ \bibinfo {author} {\bibfnamefont {J.~M.}\ \bibnamefont
  {Kosterlitz}},\ }\bibfield  {title} {\bibinfo {title} {Universal jump in the
  superfluid density of two-dimensional superfluids},\ }\href
  {https://doi.org/10.1103/PhysRevLett.39.1201} {\bibfield  {journal} {\bibinfo
   {journal} {Phys. Rev. Lett.}\ }\textbf {\bibinfo {volume} {39}},\ \bibinfo
  {pages} {1201} (\bibinfo {year} {1977})}\BibitemShut {NoStop}%
\bibitem [{\citenamefont {Peotta}\ and\ \citenamefont
  {T{\"o}rm{\"a}}(2015)}]{Peotta2015}%
  \BibitemOpen
  \bibfield  {author} {\bibinfo {author} {\bibfnamefont {S.}~\bibnamefont
  {Peotta}}\ and\ \bibinfo {author} {\bibfnamefont {P.}~\bibnamefont
  {T{\"o}rm{\"a}}},\ }\bibfield  {title} {\bibinfo {title} {Superfluidity in
  topologically nontrivial flat bands},\ }\href
  {https://doi.org/10.1038/ncomms9944} {\bibfield  {journal} {\bibinfo
  {journal} {Nature Communications}\ }\textbf {\bibinfo {volume} {6}},\
  \bibinfo {pages} {8944} (\bibinfo {year} {2015})}\BibitemShut {NoStop}%
\bibitem [{\citenamefont {Huhtinen}\ \emph {et~al.}(2022)\citenamefont
  {Huhtinen}, \citenamefont {Herzog-Arbeitman}, \citenamefont {Chew},
  \citenamefont {Bernevig},\ and\ \citenamefont {T\"orm\"a}}]{Huhtinen2022}%
  \BibitemOpen
  \bibfield  {author} {\bibinfo {author} {\bibfnamefont {K.-E.}\ \bibnamefont
  {Huhtinen}}, \bibinfo {author} {\bibfnamefont {J.}~\bibnamefont
  {Herzog-Arbeitman}}, \bibinfo {author} {\bibfnamefont {A.}~\bibnamefont
  {Chew}}, \bibinfo {author} {\bibfnamefont {B.~A.}\ \bibnamefont {Bernevig}},\
  and\ \bibinfo {author} {\bibfnamefont {P.}~\bibnamefont {T\"orm\"a}},\
  }\bibfield  {title} {\bibinfo {title} {Revisiting flat band
  superconductivity: Dependence on minimal quantum metric and band touchings},\
  }\href {https://doi.org/10.1103/PhysRevB.106.014518} {\bibfield  {journal}
  {\bibinfo  {journal} {Phys. Rev. B}\ }\textbf {\bibinfo {volume} {106}},\
  \bibinfo {pages} {014518} (\bibinfo {year} {2022})}\BibitemShut {NoStop}%
\bibitem [{\citenamefont {Tam}\ and\ \citenamefont
  {Peotta}(2024)}]{Peotta2024}%
  \BibitemOpen
  \bibfield  {author} {\bibinfo {author} {\bibfnamefont {M.}~\bibnamefont
  {Tam}}\ and\ \bibinfo {author} {\bibfnamefont {S.}~\bibnamefont {Peotta}},\
  }\bibfield  {title} {\bibinfo {title} {Geometry-independent superfluid weight
  in multiorbital lattices from the generalized random phase approximation},\
  }\href {https://doi.org/10.1103/PhysRevResearch.6.013256} {\bibfield
  {journal} {\bibinfo  {journal} {Phys. Rev. Res.}\ }\textbf {\bibinfo {volume}
  {6}},\ \bibinfo {pages} {013256} (\bibinfo {year} {2024})}\BibitemShut
  {NoStop}%
\bibitem [{\citenamefont {Peri}\ \emph {et~al.}(2021)\citenamefont {Peri},
  \citenamefont {Song}, \citenamefont {Bernevig},\ and\ \citenamefont
  {Huber}}]{Peri2021}%
  \BibitemOpen
  \bibfield  {author} {\bibinfo {author} {\bibfnamefont {V.}~\bibnamefont
  {Peri}}, \bibinfo {author} {\bibfnamefont {Z.-D.}\ \bibnamefont {Song}},
  \bibinfo {author} {\bibfnamefont {B.~A.}\ \bibnamefont {Bernevig}},\ and\
  \bibinfo {author} {\bibfnamefont {S.~D.}\ \bibnamefont {Huber}},\ }\bibfield
  {title} {\bibinfo {title} {Fragile topology and flat-band superconductivity
  in the strong-coupling regime},\ }\href
  {https://doi.org/10.1103/PhysRevLett.126.027002} {\bibfield  {journal}
  {\bibinfo  {journal} {Physical Review Letters}\ }\textbf {\bibinfo {volume}
  {126}},\ \bibinfo {pages} {027002} (\bibinfo {year} {2021})}\BibitemShut
  {NoStop}%
\bibitem [{\citenamefont {Herzog-Arbeitman}\ \emph {et~al.}(2022)\citenamefont
  {Herzog-Arbeitman}, \citenamefont {Peri}, \citenamefont {Schindler},
  \citenamefont {Huber},\ and\ \citenamefont
  {Bernevig}}]{Herzog-Arbeitman2022}%
  \BibitemOpen
  \bibfield  {author} {\bibinfo {author} {\bibfnamefont {J.}~\bibnamefont
  {Herzog-Arbeitman}}, \bibinfo {author} {\bibfnamefont {V.}~\bibnamefont
  {Peri}}, \bibinfo {author} {\bibfnamefont {F.}~\bibnamefont {Schindler}},
  \bibinfo {author} {\bibfnamefont {S.~D.}\ \bibnamefont {Huber}},\ and\
  \bibinfo {author} {\bibfnamefont {B.~A.}\ \bibnamefont {Bernevig}},\
  }\bibfield  {title} {\bibinfo {title} {Superfluid weight bounds from symmetry
  and quantum geometry in flat bands},\ }\href
  {https://doi.org/10.1103/PhysRevLett.128.087002} {\bibfield  {journal}
  {\bibinfo  {journal} {Physical Review Letters}\ }\textbf {\bibinfo {volume}
  {128}},\ \bibinfo {pages} {087002} (\bibinfo {year} {2022})}\BibitemShut
  {NoStop}%
\bibitem [{\citenamefont {Liang}\ \emph {et~al.}(2017)\citenamefont {Liang},
  \citenamefont {Vanhala}, \citenamefont {Peotta}, \citenamefont {Siro},
  \citenamefont {Harju},\ and\ \citenamefont {T{\"o}rm{\"a}}}]{Liang2017}%
  \BibitemOpen
  \bibfield  {author} {\bibinfo {author} {\bibfnamefont {L.}~\bibnamefont
  {Liang}}, \bibinfo {author} {\bibfnamefont {T.~I.}\ \bibnamefont {Vanhala}},
  \bibinfo {author} {\bibfnamefont {S.}~\bibnamefont {Peotta}}, \bibinfo
  {author} {\bibfnamefont {T.}~\bibnamefont {Siro}}, \bibinfo {author}
  {\bibfnamefont {A.}~\bibnamefont {Harju}},\ and\ \bibinfo {author}
  {\bibfnamefont {P.}~\bibnamefont {T{\"o}rm{\"a}}},\ }\bibfield  {title}
  {\bibinfo {title} {Band geometry, {Berry} curvature, and superfluid weight},\
  }\href {https://doi.org/10.1103/PhysRevB.95.024515} {\bibfield  {journal}
  {\bibinfo  {journal} {Phys. Rev. B}\ }\textbf {\bibinfo {volume} {95}},\
  \bibinfo {pages} {024515} (\bibinfo {year} {2017})}\BibitemShut {NoStop}%
\bibitem [{\citenamefont {Penttil{\"a}}\ \emph {et~al.}(2025)\citenamefont
  {Penttil{\"a}}, \citenamefont {Huhtinen},\ and\ \citenamefont
  {T{\"o}rm{\"a}}}]{Penttila2025}%
  \BibitemOpen
  \bibfield  {author} {\bibinfo {author} {\bibfnamefont {R.~P.~S.}\
  \bibnamefont {Penttil{\"a}}}, \bibinfo {author} {\bibfnamefont {K.-E.}\
  \bibnamefont {Huhtinen}},\ and\ \bibinfo {author} {\bibfnamefont
  {P.}~\bibnamefont {T{\"o}rm{\"a}}},\ }\bibfield  {title} {\bibinfo {title}
  {Flat-band ratio and quantum metric in the superconductivity of modified
  {Lieb} lattices},\ }\href {https://doi.org/10.1038/s42005-025-01964-y}
  {\bibfield  {journal} {\bibinfo  {journal} {Communications Physics}\ }\textbf
  {\bibinfo {volume} {8}},\ \bibinfo {pages} {50} (\bibinfo {year}
  {2025})}\BibitemShut {NoStop}%
\bibitem [{\citenamefont {Tovmasyan}\ \emph {et~al.}(2018)\citenamefont
  {Tovmasyan}, \citenamefont {Peotta}, \citenamefont {Liang}, \citenamefont
  {T\"orm\"a},\ and\ \citenamefont {Huber}}]{Tovmasyan18a}%
  \BibitemOpen
  \bibfield  {author} {\bibinfo {author} {\bibfnamefont {M.}~\bibnamefont
  {Tovmasyan}}, \bibinfo {author} {\bibfnamefont {S.}~\bibnamefont {Peotta}},
  \bibinfo {author} {\bibfnamefont {L.}~\bibnamefont {Liang}}, \bibinfo
  {author} {\bibfnamefont {P.}~\bibnamefont {T\"orm\"a}},\ and\ \bibinfo
  {author} {\bibfnamefont {S.~D.}\ \bibnamefont {Huber}},\ }\bibfield  {title}
  {\bibinfo {title} {Preformed pairs in flat {Bloch} bands},\ }\href
  {https://doi.org/10.1103/PhysRevB.98.134513} {\bibfield  {journal} {\bibinfo
  {journal} {Phys. Rev. B}\ }\textbf {\bibinfo {volume} {98}},\ \bibinfo
  {pages} {134513} (\bibinfo {year} {2018})}\BibitemShut {NoStop}%
\bibitem [{\citenamefont {Julku}\ \emph {et~al.}(2016)\citenamefont {Julku},
  \citenamefont {Peotta}, \citenamefont {Vanhala}, \citenamefont {Kim},\ and\
  \citenamefont {T\"orm\"a}}]{Julku2016}%
  \BibitemOpen
  \bibfield  {author} {\bibinfo {author} {\bibfnamefont {A.}~\bibnamefont
  {Julku}}, \bibinfo {author} {\bibfnamefont {S.}~\bibnamefont {Peotta}},
  \bibinfo {author} {\bibfnamefont {T.~I.}\ \bibnamefont {Vanhala}}, \bibinfo
  {author} {\bibfnamefont {D.-H.}\ \bibnamefont {Kim}},\ and\ \bibinfo {author}
  {\bibfnamefont {P.}~\bibnamefont {T\"orm\"a}},\ }\bibfield  {title} {\bibinfo
  {title} {Geometric origin of superfluidity in the {Lieb}-lattice flat band},\
  }\href {https://doi.org/10.1103/PhysRevLett.117.045303} {\bibfield  {journal}
  {\bibinfo  {journal} {Phys. Rev. Lett.}\ }\textbf {\bibinfo {volume} {117}},\
  \bibinfo {pages} {045303} (\bibinfo {year} {2016})}\BibitemShut {NoStop}%
\bibitem [{\citenamefont {Tovmasyan}\ \emph {et~al.}(2016)\citenamefont
  {Tovmasyan}, \citenamefont {Peotta}, \citenamefont {T{\"o}rm{\"a}},\ and\
  \citenamefont {Huber}}]{Tovmasyan16}%
  \BibitemOpen
  \bibfield  {author} {\bibinfo {author} {\bibfnamefont {M.}~\bibnamefont
  {Tovmasyan}}, \bibinfo {author} {\bibfnamefont {S.}~\bibnamefont {Peotta}},
  \bibinfo {author} {\bibfnamefont {P.}~\bibnamefont {T{\"o}rm{\"a}}},\ and\
  \bibinfo {author} {\bibfnamefont {S.~D.}\ \bibnamefont {Huber}},\ }\bibfield
  {title} {\bibinfo {title} {Effective theory and emergent {$\text{SU}(2)$}
  symmetry in the flat bands of attractive {Hubbard} models},\ }\href
  {https://doi.org/10.1103/PhysRevB.94.245149} {\bibfield  {journal} {\bibinfo
  {journal} {Physical Review B}\ }\textbf {\bibinfo {volume} {94}},\ \bibinfo
  {pages} {245149} (\bibinfo {year} {2016})}\BibitemShut {NoStop}%
\bibitem [{\citenamefont {{Herzog-Arbeitman}}\ \emph
  {et~al.}(2022)\citenamefont {{Herzog-Arbeitman}}, \citenamefont {{Chew}},
  \citenamefont {{Huhtinen}}, \citenamefont {{T{\"o}rm{\"a}}},\ and\
  \citenamefont {{Bernevig}}}]{Herzog-Arbeitman2022b}%
  \BibitemOpen
  \bibfield  {author} {\bibinfo {author} {\bibfnamefont {J.}~\bibnamefont
  {{Herzog-Arbeitman}}}, \bibinfo {author} {\bibfnamefont {A.}~\bibnamefont
  {{Chew}}}, \bibinfo {author} {\bibfnamefont {K.-E.}\ \bibnamefont
  {{Huhtinen}}}, \bibinfo {author} {\bibfnamefont {P.}~\bibnamefont
  {{T{\"o}rm{\"a}}}},\ and\ \bibinfo {author} {\bibfnamefont {B.~A.}\
  \bibnamefont {{Bernevig}}},\ }\bibfield  {title} {\bibinfo {title}
  {{Many-Body Superconductivity in Topological Flat Bands}},\ }\href
  {https://doi.org/10.48550/arXiv.2209.00007} {\bibfield  {journal} {\bibinfo
  {journal} {arXiv e-prints}\ ,\ \bibinfo {eid} {arXiv:2209.00007}} (\bibinfo
  {year} {2022})},\ \Eprint {https://arxiv.org/abs/2209.00007}
  {arXiv:2209.00007 [cond-mat.str-el]} \BibitemShut {NoStop}%
\bibitem [{\citenamefont {T\"orm\"a}\ \emph {et~al.}(2018)\citenamefont
  {T\"orm\"a}, \citenamefont {Liang},\ and\ \citenamefont
  {Peotta}}]{Torma2018}%
  \BibitemOpen
  \bibfield  {author} {\bibinfo {author} {\bibfnamefont {P.}~\bibnamefont
  {T\"orm\"a}}, \bibinfo {author} {\bibfnamefont {L.}~\bibnamefont {Liang}},\
  and\ \bibinfo {author} {\bibfnamefont {S.}~\bibnamefont {Peotta}},\
  }\bibfield  {title} {\bibinfo {title} {Quantum metric and effective mass of a
  two-body bound state in a flat band},\ }\href
  {https://doi.org/10.1103/PhysRevB.98.220511} {\bibfield  {journal} {\bibinfo
  {journal} {Phys. Rev. B}\ }\textbf {\bibinfo {volume} {98}},\ \bibinfo
  {pages} {220511} (\bibinfo {year} {2018})}\BibitemShut {NoStop}%
\bibitem [{\citenamefont {Iskin}(2021)}]{Iskin2021}%
  \BibitemOpen
  \bibfield  {author} {\bibinfo {author} {\bibfnamefont {M.}~\bibnamefont
  {Iskin}},\ }\bibfield  {title} {\bibinfo {title} {Two-body problem in a
  multiband lattice and the role of quantum geometry},\ }\href
  {https://doi.org/10.1103/PhysRevA.103.053311} {\bibfield  {journal} {\bibinfo
   {journal} {Phys. Rev. A}\ }\textbf {\bibinfo {volume} {103}},\ \bibinfo
  {pages} {053311} (\bibinfo {year} {2021})}\BibitemShut {NoStop}%
\bibitem [{\citenamefont {Iskin}(2022)}]{Iskin2022}%
  \BibitemOpen
  \bibfield  {author} {\bibinfo {author} {\bibfnamefont {M.}~\bibnamefont
  {Iskin}},\ }\bibfield  {title} {\bibinfo {title} {Effective-mass tensor of
  the two-body bound states and the quantum-metric tensor of the underlying
  {Bloch} states in multiband lattices},\ }\href
  {https://doi.org/10.1103/PhysRevA.105.023312} {\bibfield  {journal} {\bibinfo
   {journal} {Phys. Rev. A}\ }\textbf {\bibinfo {volume} {105}},\ \bibinfo
  {pages} {023312} (\bibinfo {year} {2022})}\BibitemShut {NoStop}%
\bibitem [{\citenamefont {Roy}(2014)}]{Roy2014}%
  \BibitemOpen
  \bibfield  {author} {\bibinfo {author} {\bibfnamefont {R.}~\bibnamefont
  {Roy}},\ }\bibfield  {title} {\bibinfo {title} {Band geometry of fractional
  topological insulators},\ }\href {https://doi.org/10.1103/PhysRevB.90.165139}
  {\bibfield  {journal} {\bibinfo  {journal} {Phys. Rev. B}\ }\textbf {\bibinfo
  {volume} {90}},\ \bibinfo {pages} {165139} (\bibinfo {year}
  {2014})}\BibitemShut {NoStop}%
\bibitem [{\citenamefont {Ozawa}\ and\ \citenamefont {Mera}(2021)}]{Ozawa2021}%
  \BibitemOpen
  \bibfield  {author} {\bibinfo {author} {\bibfnamefont {T.}~\bibnamefont
  {Ozawa}}\ and\ \bibinfo {author} {\bibfnamefont {B.}~\bibnamefont {Mera}},\
  }\bibfield  {title} {\bibinfo {title} {Relations between topology and the
  quantum metric for {Chern} insulators},\ }\href
  {https://doi.org/10.1103/PhysRevB.104.045103} {\bibfield  {journal} {\bibinfo
   {journal} {Phys. Rev. B}\ }\textbf {\bibinfo {volume} {104}},\ \bibinfo
  {pages} {045103} (\bibinfo {year} {2021})}\BibitemShut {NoStop}%
\bibitem [{\citenamefont {Mera}\ \emph {et~al.}(2022)\citenamefont {Mera},
  \citenamefont {Zhang},\ and\ \citenamefont {Goldman}}]{Mera2022}%
  \BibitemOpen
  \bibfield  {author} {\bibinfo {author} {\bibfnamefont {B.}~\bibnamefont
  {Mera}}, \bibinfo {author} {\bibfnamefont {A.}~\bibnamefont {Zhang}},\ and\
  \bibinfo {author} {\bibfnamefont {N.}~\bibnamefont {Goldman}},\ }\bibfield
  {title} {\bibinfo {title} {{Relating the topology of Dirac Hamiltonians to
  quantum geometry: When the quantum metric dictates Chern numbers and winding
  numbers}},\ }\href {https://doi.org/10.21468/SciPostPhys.12.1.018} {\bibfield
   {journal} {\bibinfo  {journal} {SciPost Phys.}\ }\textbf {\bibinfo {volume}
  {12}},\ \bibinfo {pages} {018} (\bibinfo {year} {2022})}\BibitemShut
  {NoStop}%
\bibitem [{\citenamefont {Xie}\ \emph {et~al.}(2020)\citenamefont {Xie},
  \citenamefont {Song}, \citenamefont {Lian},\ and\ \citenamefont
  {Bernevig}}]{Xie2020}%
  \BibitemOpen
  \bibfield  {author} {\bibinfo {author} {\bibfnamefont {F.}~\bibnamefont
  {Xie}}, \bibinfo {author} {\bibfnamefont {Z.}~\bibnamefont {Song}}, \bibinfo
  {author} {\bibfnamefont {B.}~\bibnamefont {Lian}},\ and\ \bibinfo {author}
  {\bibfnamefont {B.~A.}\ \bibnamefont {Bernevig}},\ }\bibfield  {title}
  {\bibinfo {title} {Topology-bounded superfluid weight in twisted bilayer
  graphene},\ }\href {https://doi.org/10.1103/PhysRevLett.124.167002}
  {\bibfield  {journal} {\bibinfo  {journal} {Physical Review Letters}\
  }\textbf {\bibinfo {volume} {124}},\ \bibinfo {pages} {167002} (\bibinfo
  {year} {2020})}\BibitemShut {NoStop}%
\bibitem [{\citenamefont {Jankowski}\ \emph {et~al.}(2024)\citenamefont
  {Jankowski}, \citenamefont {Morris}, \citenamefont {Davoyan}, \citenamefont
  {Bouhon}, \citenamefont {\"{U}nal},\ and\ \citenamefont
  {Slager}}]{Jankowski2024}%
  \BibitemOpen
  \bibfield  {author} {\bibinfo {author} {\bibfnamefont {W.~J.}\ \bibnamefont
  {Jankowski}}, \bibinfo {author} {\bibfnamefont {A.~S.}\ \bibnamefont
  {Morris}}, \bibinfo {author} {\bibfnamefont {Z.}~\bibnamefont {Davoyan}},
  \bibinfo {author} {\bibfnamefont {A.}~\bibnamefont {Bouhon}}, \bibinfo
  {author} {\bibfnamefont {F.~N.}\ \bibnamefont {\"{U}nal}},\ and\ \bibinfo
  {author} {\bibfnamefont {R.-J.}\ \bibnamefont {Slager}},\ }\bibfield  {title}
  {\bibinfo {title} {Non-abelian {Hopf-Euler} insulators},\ }\bibfield
  {journal} {\bibinfo  {journal} {Physical Review B}\ }\textbf {\bibinfo
  {volume} {110}},\ \href {https://doi.org/10.1103/physrevb.110.075135}
  {10.1103/physrevb.110.075135} (\bibinfo {year} {2024})\BibitemShut {NoStop}%
\bibitem [{\citenamefont {Yu}\ \emph {et~al.}(2025{\natexlab{a}})\citenamefont
  {Yu}, \citenamefont {Herzog-Arbeitman},\ and\ \citenamefont
  {Bernevig}}]{Yu2025}%
  \BibitemOpen
  \bibfield  {author} {\bibinfo {author} {\bibfnamefont {J.}~\bibnamefont
  {Yu}}, \bibinfo {author} {\bibfnamefont {J.}~\bibnamefont
  {Herzog-Arbeitman}},\ and\ \bibinfo {author} {\bibfnamefont {B.~A.}\
  \bibnamefont {Bernevig}},\ }\bibfield  {title} {\bibinfo {title} {Universal
  {Wilson} loop bound of quantum geometry},\ }\href
  {https://doi.org/10.1103/mp2c-zzkt} {\bibfield  {journal} {\bibinfo
  {journal} {Phys. Rev. Lett.}\ }\textbf {\bibinfo {volume} {135}},\ \bibinfo
  {pages} {086401} (\bibinfo {year} {2025}{\natexlab{a}})}\BibitemShut
  {NoStop}%
\bibitem [{Note1()}]{Note1}%
  \BibitemOpen
  \bibinfo {note} {The superfluid weight tensor is of the form $D\propto
  \protect \mathbb {1}$ when the model is sufficiently symmetric, with for
  instance $C_3$ or $C_4$ rotational symmetry. We refer to these models as
  isotropic.}\BibitemShut {Stop}%
\bibitem [{\citenamefont {Nelson}\ \emph {et~al.}(2021)\citenamefont {Nelson},
  \citenamefont {Neupert}, \citenamefont {Bzdu{\v s}ek},\ and\ \citenamefont
  {Alexandradinata}}]{Nelson2021}%
  \BibitemOpen
  \bibfield  {author} {\bibinfo {author} {\bibfnamefont {A.}~\bibnamefont
  {Nelson}}, \bibinfo {author} {\bibfnamefont {T.}~\bibnamefont {Neupert}},
  \bibinfo {author} {\bibfnamefont {T.}~\bibnamefont {Bzdu{\v s}ek}},\ and\
  \bibinfo {author} {\bibfnamefont {A.}~\bibnamefont {Alexandradinata}},\
  }\bibfield  {title} {\bibinfo {title} {Multicellularity of delicate
  topological insulators},\ }\href
  {https://doi.org/10.1103/PhysRevLett.126.216404} {\bibfield  {journal}
  {\bibinfo  {journal} {Phys. Rev. Lett.}\ }\textbf {\bibinfo {volume} {126}},\
  \bibinfo {pages} {216404} (\bibinfo {year} {2021})}\BibitemShut {NoStop}%
\bibitem [{\citenamefont {Nelson}\ \emph {et~al.}(2022)\citenamefont {Nelson},
  \citenamefont {Neupert}, \citenamefont {Alexandradinata},\ and\ \citenamefont
  {Bzdu{\v s}ek}}]{Nelson2022}%
  \BibitemOpen
  \bibfield  {author} {\bibinfo {author} {\bibfnamefont {A.}~\bibnamefont
  {Nelson}}, \bibinfo {author} {\bibfnamefont {T.}~\bibnamefont {Neupert}},
  \bibinfo {author} {\bibfnamefont {A.}~\bibnamefont {Alexandradinata}},\ and\
  \bibinfo {author} {\bibfnamefont {T.}~\bibnamefont {Bzdu{\v s}ek}},\
  }\bibfield  {title} {\bibinfo {title} {Delicate topology protected by
  rotation symmetry: Crystalline {Hopf} insulators and beyond},\ }\href
  {https://doi.org/10.1103/PhysRevB.106.075124} {\bibfield  {journal} {\bibinfo
   {journal} {Physical Review B}\ }\textbf {\bibinfo {volume} {106}},\ \bibinfo
  {pages} {075124} (\bibinfo {year} {2022})}\BibitemShut {NoStop}%
\bibitem [{\citenamefont {Moore}\ \emph {et~al.}(2008)\citenamefont {Moore},
  \citenamefont {Ran},\ and\ \citenamefont {Wen}}]{Moore2008}%
  \BibitemOpen
  \bibfield  {author} {\bibinfo {author} {\bibfnamefont {J.~E.}\ \bibnamefont
  {Moore}}, \bibinfo {author} {\bibfnamefont {Y.}~\bibnamefont {Ran}},\ and\
  \bibinfo {author} {\bibfnamefont {X.-G.}\ \bibnamefont {Wen}},\ }\bibfield
  {title} {\bibinfo {title} {Topological surface states in three-dimensional
  magnetic insulators},\ }\href
  {https://doi.org/10.1103/PhysRevLett.101.186805} {\bibfield  {journal}
  {\bibinfo  {journal} {Physical Review Letters}\ }\textbf {\bibinfo {volume}
  {101}},\ \bibinfo {pages} {186805} (\bibinfo {year} {2008})}\BibitemShut
  {NoStop}%
\bibitem [{\citenamefont {Liu}\ \emph {et~al.}(2017)\citenamefont {Liu},
  \citenamefont {Vafa},\ and\ \citenamefont {Xu}}]{Liu2017}%
  \BibitemOpen
  \bibfield  {author} {\bibinfo {author} {\bibfnamefont {C.}~\bibnamefont
  {Liu}}, \bibinfo {author} {\bibfnamefont {F.}~\bibnamefont {Vafa}},\ and\
  \bibinfo {author} {\bibfnamefont {C.}~\bibnamefont {Xu}},\ }\bibfield
  {title} {\bibinfo {title} {Symmetry-protected topological {Hopf} insulator
  and its generalizations},\ }\bibfield  {journal} {\bibinfo  {journal}
  {Physical Review B}\ }\textbf {\bibinfo {volume} {95}},\ \href
  {https://doi.org/10.1103/physrevb.95.161116} {10.1103/physrevb.95.161116}
  (\bibinfo {year} {2017})\BibitemShut {NoStop}%
\bibitem [{\citenamefont {Bradlyn}\ \emph {et~al.}(2017)\citenamefont
  {Bradlyn}, \citenamefont {Elcoro}, \citenamefont {Cano}, \citenamefont
  {Vergniory}, \citenamefont {Wang}, \citenamefont {Felser}, \citenamefont
  {Aroyo},\ and\ \citenamefont {Bernevig}}]{Bradlyn17}%
  \BibitemOpen
  \bibfield  {author} {\bibinfo {author} {\bibfnamefont {B.}~\bibnamefont
  {Bradlyn}}, \bibinfo {author} {\bibfnamefont {L.}~\bibnamefont {Elcoro}},
  \bibinfo {author} {\bibfnamefont {J.}~\bibnamefont {Cano}}, \bibinfo {author}
  {\bibfnamefont {M.~G.}\ \bibnamefont {Vergniory}}, \bibinfo {author}
  {\bibfnamefont {Z.}~\bibnamefont {Wang}}, \bibinfo {author} {\bibfnamefont
  {C.}~\bibnamefont {Felser}}, \bibinfo {author} {\bibfnamefont {M.~I.}\
  \bibnamefont {Aroyo}},\ and\ \bibinfo {author} {\bibfnamefont {B.~A.}\
  \bibnamefont {Bernevig}},\ }\bibfield  {title} {\bibinfo {title} {Topological
  quantum chemistry},\ }\href {https://doi.org/10.1038/nature23268} {\bibfield
  {journal} {\bibinfo  {journal} {Nature}\ }\textbf {\bibinfo {volume} {547}},\
  \bibinfo {pages} {298} (\bibinfo {year} {2017})}\BibitemShut {NoStop}%
\bibitem [{\citenamefont {Po}\ \emph {et~al.}(2018)\citenamefont {Po},
  \citenamefont {Watanabe},\ and\ \citenamefont {Vishwanath}}]{Po2018}%
  \BibitemOpen
  \bibfield  {author} {\bibinfo {author} {\bibfnamefont {H.~C.}\ \bibnamefont
  {Po}}, \bibinfo {author} {\bibfnamefont {H.}~\bibnamefont {Watanabe}},\ and\
  \bibinfo {author} {\bibfnamefont {A.}~\bibnamefont {Vishwanath}},\ }\bibfield
   {title} {\bibinfo {title} {Fragile topology and {Wannier} obstructions},\
  }\href {https://doi.org/10.1103/PhysRevLett.121.126402} {\bibfield  {journal}
  {\bibinfo  {journal} {Phys. Rev. Lett.}\ }\textbf {\bibinfo {volume} {121}},\
  \bibinfo {pages} {126402} (\bibinfo {year} {2018})}\BibitemShut {NoStop}%
\bibitem [{\citenamefont {Chen}\ \emph {et~al.}(2024)\citenamefont {Chen},
  \citenamefont {Lin},\ and\ \citenamefont {Kao}}]{Chen2024}%
  \BibitemOpen
  \bibfield  {author} {\bibinfo {author} {\bibfnamefont {Y.-C.}\ \bibnamefont
  {Chen}}, \bibinfo {author} {\bibfnamefont {Y.-P.}\ \bibnamefont {Lin}},\ and\
  \bibinfo {author} {\bibfnamefont {Y.-J.}\ \bibnamefont {Kao}},\ }\bibfield
  {title} {\bibinfo {title} {Chern dartboard insulator: sub-{Brillouin} zone
  topology and skyrmion multipoles},\ }\href
  {https://doi.org/10.1038/s42005-023-01502-8} {\bibfield  {journal} {\bibinfo
  {journal} {Commun. Phys.}\ }\textbf {\bibinfo {volume} {7}},\ \bibinfo
  {pages} {32} (\bibinfo {year} {2024})}\BibitemShut {NoStop}%
\bibitem [{sup()}]{supplemental}%
  \BibitemOpen
  \href@noop {} {}\bibinfo {note} {See supplemental material for the dependence
  of the Chern number on the basis, expressions for the superfluid weight in
  systems with uniform pairing, and model Hamiltonians.}\BibitemShut {Stop}%
\bibitem [{\citenamefont {Chan}\ \emph {et~al.}(2022)\citenamefont {Chan},
  \citenamefont {Gr\'emaud},\ and\ \citenamefont {Batrouni}}]{Chan2022}%
  \BibitemOpen
  \bibfield  {author} {\bibinfo {author} {\bibfnamefont {S.~M.}\ \bibnamefont
  {Chan}}, \bibinfo {author} {\bibfnamefont {B.}~\bibnamefont {Gr\'emaud}},\
  and\ \bibinfo {author} {\bibfnamefont {G.~G.}\ \bibnamefont {Batrouni}},\
  }\bibfield  {title} {\bibinfo {title} {Pairing and superconductivity in
  quasi-one-dimensional flat-band systems: Creutz and sawtooth lattices},\
  }\href {https://doi.org/10.1103/PhysRevB.105.024502} {\bibfield  {journal}
  {\bibinfo  {journal} {Phys. Rev. B}\ }\textbf {\bibinfo {volume} {105}},\
  \bibinfo {pages} {024502} (\bibinfo {year} {2022})}\BibitemShut {NoStop}%
\bibitem [{\citenamefont {{Thumin}}\ and\ \citenamefont
  {{Bouzerar}}(2025)}]{Thumin2025}%
  \BibitemOpen
  \bibfield  {author} {\bibinfo {author} {\bibfnamefont {M.}~\bibnamefont
  {{Thumin}}}\ and\ \bibinfo {author} {\bibfnamefont {G.}~\bibnamefont
  {{Bouzerar}}},\ }\bibfield  {title} {\bibinfo {title} {{Crossing over from
  flat band superconductivity to conventional superconductivity}},\ }\href
  {https://doi.org/10.48550/arXiv.2507.07701} {\bibfield  {journal} {\bibinfo
  {journal} {arXiv e-prints}\ ,\ \bibinfo {eid} {arXiv:2507.07701}} (\bibinfo
  {year} {2025})},\ \Eprint {https://arxiv.org/abs/2507.07701}
  {arXiv:2507.07701 [cond-mat.supr-con]} \BibitemShut {NoStop}%
\bibitem [{\citenamefont {Vaidya}\ \emph {et~al.}(2025)\citenamefont {Vaidya},
  \citenamefont {Fonseca}, \citenamefont {Hirsbrunner}, \citenamefont
  {Hughes},\ and\ \citenamefont {Solja\ifmmode \check{c}\else
  \v{c}\fi{}i\ifmmode~\acute{c}\else \'{c}\fi{}}}]{Vaidya2025}%
  \BibitemOpen
  \bibfield  {author} {\bibinfo {author} {\bibfnamefont {S.}~\bibnamefont
  {Vaidya}}, \bibinfo {author} {\bibfnamefont {A.~G.}\ \bibnamefont {Fonseca}},
  \bibinfo {author} {\bibfnamefont {M.~R.}\ \bibnamefont {Hirsbrunner}},
  \bibinfo {author} {\bibfnamefont {T.~L.}\ \bibnamefont {Hughes}},\ and\
  \bibinfo {author} {\bibfnamefont {M.}~\bibnamefont {Solja\ifmmode
  \check{c}\else \v{c}\fi{}i\ifmmode~\acute{c}\else \'{c}\fi{}}},\ }\bibfield
  {title} {\bibinfo {title} {Quantized crystalline-electromagnetic responses in
  insulators},\ }\href {https://doi.org/10.1103/5nsr-rw4d} {\bibfield
  {journal} {\bibinfo  {journal} {Phys. Rev. Lett.}\ }\textbf {\bibinfo
  {volume} {135}},\ \bibinfo {pages} {256602} (\bibinfo {year}
  {2025})}\BibitemShut {NoStop}%
\bibitem [{\citenamefont {Zhu}\ and\ \citenamefont
  {Alexandradinata}(2024)}]{Zhu2024}%
  \BibitemOpen
  \bibfield  {author} {\bibinfo {author} {\bibfnamefont {P.}~\bibnamefont
  {Zhu}}\ and\ \bibinfo {author} {\bibfnamefont {A.}~\bibnamefont
  {Alexandradinata}},\ }\bibfield  {title} {\bibinfo {title} {Anomalous shift
  and optical vorticity in the steady photovoltaic current},\ }\href@noop {}
  {\bibfield  {journal} {\bibinfo  {journal} {Phys. Rev. B}\ }\textbf {\bibinfo
  {volume} {110}},\ \bibinfo {pages} {115108} (\bibinfo {year}
  {2024})}\BibitemShut {NoStop}%
\bibitem [{\citenamefont {Rossi}(2021)}]{Rossi2021}%
  \BibitemOpen
  \bibfield  {author} {\bibinfo {author} {\bibfnamefont {E.}~\bibnamefont
  {Rossi}},\ }\bibfield  {title} {\bibinfo {title} {Quantum metric and
  correlated states in two-dimensional systems},\ }\href
  {https://doi.org/https://doi.org/10.1016/j.cossms.2021.100952} {\bibfield
  {journal} {\bibinfo  {journal} {Current Opinion in Solid State and Materials
  Science}\ }\textbf {\bibinfo {volume} {25}},\ \bibinfo {pages} {100952}
  (\bibinfo {year} {2021})}\BibitemShut {NoStop}%
\bibitem [{\citenamefont {Liu}\ \emph {et~al.}(2024)\citenamefont {Liu},
  \citenamefont {Qiang}, \citenamefont {Lu},\ and\ \citenamefont
  {Xie}}]{Liu2024}%
  \BibitemOpen
  \bibfield  {author} {\bibinfo {author} {\bibfnamefont {T.}~\bibnamefont
  {Liu}}, \bibinfo {author} {\bibfnamefont {X.-B.}\ \bibnamefont {Qiang}},
  \bibinfo {author} {\bibfnamefont {H.-Z.}\ \bibnamefont {Lu}},\ and\ \bibinfo
  {author} {\bibfnamefont {X.~C.}\ \bibnamefont {Xie}},\ }\bibfield  {title}
  {\bibinfo {title} {Quantum geometry in condensed matter},\ }\href
  {https://doi.org/10.1093/nsr/nwae334} {\bibfield  {journal} {\bibinfo
  {journal} {National Science Review}\ }\textbf {\bibinfo {volume} {12}},\
  \bibinfo {pages} {nwae334} (\bibinfo {year} {2024})},\ \Eprint
  {https://arxiv.org/abs/https://academic.oup.com/nsr/article-pdf/12/3/nwae334/59204701/nwae334.pdf}
  {https://academic.oup.com/nsr/article-pdf/12/3/nwae334/59204701/nwae334.pdf}
  \BibitemShut {NoStop}%
\bibitem [{\citenamefont {Yu}\ \emph {et~al.}(2025{\natexlab{b}})\citenamefont
  {Yu}, \citenamefont {Bernevig}, \citenamefont {Queiroz}, \citenamefont
  {Rossi}, \citenamefont {T{\"o}rm{\"a}},\ and\ \citenamefont {Yang}}]{Yu2024}%
  \BibitemOpen
  \bibfield  {author} {\bibinfo {author} {\bibfnamefont {J.}~\bibnamefont
  {Yu}}, \bibinfo {author} {\bibfnamefont {B.~A.}\ \bibnamefont {Bernevig}},
  \bibinfo {author} {\bibfnamefont {R.}~\bibnamefont {Queiroz}}, \bibinfo
  {author} {\bibfnamefont {E.}~\bibnamefont {Rossi}}, \bibinfo {author}
  {\bibfnamefont {P.}~\bibnamefont {T{\"o}rm{\"a}}},\ and\ \bibinfo {author}
  {\bibfnamefont {B.-J.}\ \bibnamefont {Yang}},\ }\bibfield  {title} {\bibinfo
  {title} {Quantum geometry in quantum materials},\ }\href
  {https://doi.org/10.1038/s41535-025-00801-3} {\bibfield  {journal} {\bibinfo
  {journal} {npj Quantum Materials}\ }\textbf {\bibinfo {volume} {10}},\
  \bibinfo {pages} {101} (\bibinfo {year} {2025}{\natexlab{b}})}\BibitemShut
  {NoStop}%
\bibitem [{\citenamefont {T\"orm\"a}(2023)}]{Torma2023}%
  \BibitemOpen
  \bibfield  {author} {\bibinfo {author} {\bibfnamefont {P.}~\bibnamefont
  {T\"orm\"a}},\ }\bibfield  {title} {\bibinfo {title} {Essay: Where can
  quantum geometry lead us?},\ }\href
  {https://doi.org/10.1103/PhysRevLett.131.240001} {\bibfield  {journal}
  {\bibinfo  {journal} {Phys. Rev. Lett.}\ }\textbf {\bibinfo {volume} {131}},\
  \bibinfo {pages} {240001} (\bibinfo {year} {2023})}\BibitemShut {NoStop}%
\bibitem [{\citenamefont {Jiang}\ \emph {et~al.}(2025)\citenamefont {Jiang},
  \citenamefont {Holder},\ and\ \citenamefont {Yan}}]{Jiang2025}%
  \BibitemOpen
  \bibfield  {author} {\bibinfo {author} {\bibfnamefont {Y.}~\bibnamefont
  {Jiang}}, \bibinfo {author} {\bibfnamefont {T.}~\bibnamefont {Holder}},\ and\
  \bibinfo {author} {\bibfnamefont {B.}~\bibnamefont {Yan}},\ }\bibfield
  {title} {\bibinfo {title} {Revealing quantum geometry in nonlinear quantum
  materials},\ }\href {https://doi.org/10.1088/1361-6633/ade454} {\bibfield
  {journal} {\bibinfo  {journal} {Reports on Progress in Physics}\ }\textbf
  {\bibinfo {volume} {88}},\ \bibinfo {pages} {076502} (\bibinfo {year}
  {2025})}\BibitemShut {NoStop}%
\bibitem [{\citenamefont {Gao}\ \emph {et~al.}(2014)\citenamefont {Gao},
  \citenamefont {Yang},\ and\ \citenamefont {Niu}}]{Gao2014}%
  \BibitemOpen
  \bibfield  {author} {\bibinfo {author} {\bibfnamefont {Y.}~\bibnamefont
  {Gao}}, \bibinfo {author} {\bibfnamefont {S.~A.}\ \bibnamefont {Yang}},\ and\
  \bibinfo {author} {\bibfnamefont {Q.}~\bibnamefont {Niu}},\ }\bibfield
  {title} {\bibinfo {title} {Field induced positional shift of bloch electrons
  and its dynamical implications},\ }\href
  {https://doi.org/10.1103/PhysRevLett.112.166601} {\bibfield  {journal}
  {\bibinfo  {journal} {Phys. Rev. Lett.}\ }\textbf {\bibinfo {volume} {112}},\
  \bibinfo {pages} {166601} (\bibinfo {year} {2014})}\BibitemShut {NoStop}%
\bibitem [{\citenamefont {Kaplan}\ \emph {et~al.}(2024)\citenamefont {Kaplan},
  \citenamefont {Holder},\ and\ \citenamefont {Yan}}]{Kaplan2024}%
  \BibitemOpen
  \bibfield  {author} {\bibinfo {author} {\bibfnamefont {D.}~\bibnamefont
  {Kaplan}}, \bibinfo {author} {\bibfnamefont {T.}~\bibnamefont {Holder}},\
  and\ \bibinfo {author} {\bibfnamefont {B.}~\bibnamefont {Yan}},\ }\bibfield
  {title} {\bibinfo {title} {Unification of nonlinear anomalous {Hall} effect
  and nonreciprocal magnetoresistance in metals by the quantum geometry},\
  }\href {https://doi.org/10.1103/PhysRevLett.132.026301} {\bibfield  {journal}
  {\bibinfo  {journal} {Phys. Rev. Lett.}\ }\textbf {\bibinfo {volume} {132}},\
  \bibinfo {pages} {026301} (\bibinfo {year} {2024})}\BibitemShut {NoStop}%
\bibitem [{\citenamefont {Wang}\ \emph {et~al.}(2023)\citenamefont {Wang},
  \citenamefont {Kaplan}, \citenamefont {Zhang}, \citenamefont {Holder},
  \citenamefont {Cao}, \citenamefont {Wang}, \citenamefont {Zhou},
  \citenamefont {Zhou}, \citenamefont {Jiang}, \citenamefont {Zhang},
  \citenamefont {Ru}, \citenamefont {Cai}, \citenamefont {Watanabe},
  \citenamefont {Taniguchi}, \citenamefont {Yan},\ and\ \citenamefont
  {Gao}}]{Wang2023}%
  \BibitemOpen
  \bibfield  {author} {\bibinfo {author} {\bibfnamefont {N.}~\bibnamefont
  {Wang}}, \bibinfo {author} {\bibfnamefont {D.}~\bibnamefont {Kaplan}},
  \bibinfo {author} {\bibfnamefont {Z.}~\bibnamefont {Zhang}}, \bibinfo
  {author} {\bibfnamefont {T.}~\bibnamefont {Holder}}, \bibinfo {author}
  {\bibfnamefont {N.}~\bibnamefont {Cao}}, \bibinfo {author} {\bibfnamefont
  {A.}~\bibnamefont {Wang}}, \bibinfo {author} {\bibfnamefont {X.}~\bibnamefont
  {Zhou}}, \bibinfo {author} {\bibfnamefont {F.}~\bibnamefont {Zhou}}, \bibinfo
  {author} {\bibfnamefont {Z.}~\bibnamefont {Jiang}}, \bibinfo {author}
  {\bibfnamefont {C.}~\bibnamefont {Zhang}}, \bibinfo {author} {\bibfnamefont
  {S.}~\bibnamefont {Ru}}, \bibinfo {author} {\bibfnamefont {H.}~\bibnamefont
  {Cai}}, \bibinfo {author} {\bibfnamefont {K.}~\bibnamefont {Watanabe}},
  \bibinfo {author} {\bibfnamefont {T.}~\bibnamefont {Taniguchi}}, \bibinfo
  {author} {\bibfnamefont {B.}~\bibnamefont {Yan}},\ and\ \bibinfo {author}
  {\bibfnamefont {W.}~\bibnamefont {Gao}},\ }\bibfield  {title} {\bibinfo
  {title} {Quantum-metric-induced nonlinear transport in a topological
  antiferromagnet},\ }\href {https://doi.org/10.1038/s41586-023-06363-3}
  {\bibfield  {journal} {\bibinfo  {journal} {Nature}\ }\textbf {\bibinfo
  {volume} {621}},\ \bibinfo {pages} {487} (\bibinfo {year}
  {2023})}\BibitemShut {NoStop}%
\bibitem [{\citenamefont {{Ulrich}}\ \emph {et~al.}(2025)\citenamefont
  {{Ulrich}}, \citenamefont {{Mitscherling}}, \citenamefont {{Classen}},\ and\
  \citenamefont {{Schnyder}}}]{Ulrich2025}%
  \BibitemOpen
  \bibfield  {author} {\bibinfo {author} {\bibfnamefont {Y.}~\bibnamefont
  {{Ulrich}}}, \bibinfo {author} {\bibfnamefont {J.}~\bibnamefont
  {{Mitscherling}}}, \bibinfo {author} {\bibfnamefont {L.}~\bibnamefont
  {{Classen}}},\ and\ \bibinfo {author} {\bibfnamefont {A.~P.}\ \bibnamefont
  {{Schnyder}}},\ }\bibfield  {title} {\bibinfo {title} {{Quantum Geometric
  Origin of the Intrinsic Nonlinear Hall Effect}},\ }\href
  {https://doi.org/10.48550/arXiv.2506.17386} {\bibfield  {journal} {\bibinfo
  {journal} {arXiv e-prints}\ ,\ \bibinfo {eid} {arXiv:2506.17386}} (\bibinfo
  {year} {2025})},\ \Eprint {https://arxiv.org/abs/2506.17386}
  {arXiv:2506.17386 [cond-mat.mes-hall]} \BibitemShut {NoStop}%
\bibitem [{\citenamefont {Julku}\ \emph {et~al.}(2021)\citenamefont {Julku},
  \citenamefont {Bruun},\ and\ \citenamefont {T\"orm\"a}}]{Torma2021}%
  \BibitemOpen
  \bibfield  {author} {\bibinfo {author} {\bibfnamefont {A.}~\bibnamefont
  {Julku}}, \bibinfo {author} {\bibfnamefont {G.~M.}\ \bibnamefont {Bruun}},\
  and\ \bibinfo {author} {\bibfnamefont {P.}~\bibnamefont {T\"orm\"a}},\
  }\bibfield  {title} {\bibinfo {title} {Quantum geometry and flat band
  {Bose-Einstein} condensation},\ }\href
  {https://doi.org/10.1103/PhysRevLett.127.170404} {\bibfield  {journal}
  {\bibinfo  {journal} {Phys. Rev. Lett.}\ }\textbf {\bibinfo {volume} {127}},\
  \bibinfo {pages} {170404} (\bibinfo {year} {2021})}\BibitemShut {NoStop}%
\bibitem [{\citenamefont {{Lukin}}\ \emph {et~al.}(2023)\citenamefont
  {{Lukin}}, \citenamefont {{Sotnikov}},\ and\ \citenamefont
  {{Kruchkov}}}]{Lukin2023}%
  \BibitemOpen
  \bibfield  {author} {\bibinfo {author} {\bibfnamefont {I.}~\bibnamefont
  {{Lukin}}}, \bibinfo {author} {\bibfnamefont {A.}~\bibnamefont
  {{Sotnikov}}},\ and\ \bibinfo {author} {\bibfnamefont {A.}~\bibnamefont
  {{Kruchkov}}},\ }\bibfield  {title} {\bibinfo {title} {{Unconventional
  superfluidity and quantum geometry of topological bosons}},\ }\href
  {https://doi.org/10.48550/arXiv.2307.08748} {\bibfield  {journal} {\bibinfo
  {journal} {arXiv e-prints}\ ,\ \bibinfo {eid} {arXiv:2307.08748}} (\bibinfo
  {year} {2023})},\ \Eprint {https://arxiv.org/abs/2307.08748}
  {arXiv:2307.08748 [cond-mat.quant-gas]} \BibitemShut {NoStop}%
\bibitem [{\citenamefont {Iskin}(2023)}]{Iskin2023}%
  \BibitemOpen
  \bibfield  {author} {\bibinfo {author} {\bibfnamefont {M.}~\bibnamefont
  {Iskin}},\ }\bibfield  {title} {\bibinfo {title} {Quantum-geometric
  contribution to the {Bogoliubov} modes in a two-band {Bose-Einstein}
  condensate},\ }\href {https://doi.org/10.1103/PhysRevA.107.023313} {\bibfield
   {journal} {\bibinfo  {journal} {Phys. Rev. A}\ }\textbf {\bibinfo {volume}
  {107}},\ \bibinfo {pages} {023313} (\bibinfo {year} {2023})}\BibitemShut
  {NoStop}%
\bibitem [{\citenamefont {Resta}(2011)}]{Resta2011}%
  \BibitemOpen
  \bibfield  {author} {\bibinfo {author} {\bibfnamefont {R.}~\bibnamefont
  {Resta}},\ }\bibfield  {title} {\bibinfo {title} {The insulating state of
  matter: a geometrical theory},\ }\href
  {https://doi.org/10.1140/epjb/e2010-10874-4} {\bibfield  {journal} {\bibinfo
  {journal} {The European Physical Journal B}\ }\textbf {\bibinfo {volume}
  {79}},\ \bibinfo {pages} {121} (\bibinfo {year} {2011})}\BibitemShut
  {NoStop}%
\bibitem [{\citenamefont {Resta}(2018)}]{Resta2018}%
  \BibitemOpen
  \bibfield  {author} {\bibinfo {author} {\bibfnamefont {R.}~\bibnamefont
  {Resta}},\ }\bibfield  {title} {\bibinfo {title} {Drude weight and
  superconducting weight},\ }\href {https://doi.org/10.1088/1361-648X/aade19}
  {\bibfield  {journal} {\bibinfo  {journal} {Journal of Physics: Condensed
  Matter}\ }\textbf {\bibinfo {volume} {30}},\ \bibinfo {pages} {414001}
  (\bibinfo {year} {2018})}\BibitemShut {NoStop}%
\bibitem [{\citenamefont {Antebi}\ \emph {et~al.}(2024)\citenamefont {Antebi},
  \citenamefont {Mitscherling},\ and\ \citenamefont {Holder}}]{Antebi2024}%
  \BibitemOpen
  \bibfield  {author} {\bibinfo {author} {\bibfnamefont {O.}~\bibnamefont
  {Antebi}}, \bibinfo {author} {\bibfnamefont {J.}~\bibnamefont
  {Mitscherling}},\ and\ \bibinfo {author} {\bibfnamefont {T.}~\bibnamefont
  {Holder}},\ }\bibfield  {title} {\bibinfo {title} {Drude weight of an
  interacting flat-band metal},\ }\href
  {https://doi.org/10.1103/PhysRevB.110.L241111} {\bibfield  {journal}
  {\bibinfo  {journal} {Phys. Rev. B}\ }\textbf {\bibinfo {volume} {110}},\
  \bibinfo {pages} {L241111} (\bibinfo {year} {2024})}\BibitemShut {NoStop}%
\bibitem [{\citenamefont {Komissarov}\ \emph {et~al.}(2024)\citenamefont
  {Komissarov}, \citenamefont {Holder},\ and\ \citenamefont
  {Queiroz}}]{Komissarov2025}%
  \BibitemOpen
  \bibfield  {author} {\bibinfo {author} {\bibfnamefont {I.}~\bibnamefont
  {Komissarov}}, \bibinfo {author} {\bibfnamefont {T.}~\bibnamefont {Holder}},\
  and\ \bibinfo {author} {\bibfnamefont {R.}~\bibnamefont {Queiroz}},\
  }\bibfield  {title} {\bibinfo {title} {The quantum geometric origin of
  capacitance in insulators},\ }\href
  {https://doi.org/10.1038/s41467-024-48808-x} {\bibfield  {journal} {\bibinfo
  {journal} {Nature Communications}\ }\textbf {\bibinfo {volume} {15}},\
  \bibinfo {pages} {4621} (\bibinfo {year} {2024})}\BibitemShut {NoStop}%
\bibitem [{\citenamefont {{Verma}}\ and\ \citenamefont
  {{Queiroz}}(2024)}]{Verma2024}%
  \BibitemOpen
  \bibfield  {author} {\bibinfo {author} {\bibfnamefont {N.}~\bibnamefont
  {{Verma}}}\ and\ \bibinfo {author} {\bibfnamefont {R.}~\bibnamefont
  {{Queiroz}}},\ }\bibfield  {title} {\bibinfo {title} {{Instantaneous Response
  and Quantum Geometry of Insulators}},\ }\href
  {https://doi.org/10.48550/arXiv.2403.07052} {\bibfield  {journal} {\bibinfo
  {journal} {arXiv e-prints}\ ,\ \bibinfo {eid} {arXiv:2403.07052}} (\bibinfo
  {year} {2024})},\ \Eprint {https://arxiv.org/abs/2403.07052}
  {arXiv:2403.07052 [cond-mat.mes-hall]} \BibitemShut {NoStop}%
\bibitem [{\citenamefont {{Kruchkov}}\ and\ \citenamefont
  {{Ryu}}(2023)}]{Kruchkov2023}%
  \BibitemOpen
  \bibfield  {author} {\bibinfo {author} {\bibfnamefont {A.}~\bibnamefont
  {{Kruchkov}}}\ and\ \bibinfo {author} {\bibfnamefont {S.}~\bibnamefont
  {{Ryu}}},\ }\bibfield  {title} {\bibinfo {title} {{Spectral sum rules reflect
  topological and quantum-geometric invariants}},\ }\href
  {https://doi.org/10.48550/arXiv.2312.17318} {\bibfield  {journal} {\bibinfo
  {journal} {arXiv e-prints}\ ,\ \bibinfo {eid} {arXiv:2312.17318}} (\bibinfo
  {year} {2023})},\ \Eprint {https://arxiv.org/abs/2312.17318}
  {arXiv:2312.17318 [cond-mat.str-el]} \BibitemShut {NoStop}%
\bibitem [{\citenamefont {Mitscherling}\ and\ \citenamefont
  {Holder}(2022)}]{Mitscherling2022}%
  \BibitemOpen
  \bibfield  {author} {\bibinfo {author} {\bibfnamefont {J.}~\bibnamefont
  {Mitscherling}}\ and\ \bibinfo {author} {\bibfnamefont {T.}~\bibnamefont
  {Holder}},\ }\bibfield  {title} {\bibinfo {title} {Bound on resistivity in
  flat-band materials due to the quantum metric},\ }\href
  {https://doi.org/10.1103/PhysRevB.105.085154} {\bibfield  {journal} {\bibinfo
   {journal} {Phys. Rev. B}\ }\textbf {\bibinfo {volume} {105}},\ \bibinfo
  {pages} {085154} (\bibinfo {year} {2022})}\BibitemShut {NoStop}%
\bibitem [{\citenamefont {Bouzerar}(2022)}]{Bouzerar2022}%
  \BibitemOpen
  \bibfield  {author} {\bibinfo {author} {\bibfnamefont {G.}~\bibnamefont
  {Bouzerar}},\ }\bibfield  {title} {\bibinfo {title} {Giant boost of the
  quantum metric in disordered one-dimensional flat-band systems},\ }\href
  {https://doi.org/10.1103/PhysRevB.106.125125} {\bibfield  {journal} {\bibinfo
   {journal} {Phys. Rev. B}\ }\textbf {\bibinfo {volume} {106}},\ \bibinfo
  {pages} {125125} (\bibinfo {year} {2022})}\BibitemShut {NoStop}%
\bibitem [{\citenamefont {Prijon}\ \emph {et~al.}(2026)\citenamefont {Prijon},
  \citenamefont {Huber},\ and\ \citenamefont {Huhtinen}}]{data}%
  \BibitemOpen
  \bibfield  {author} {\bibinfo {author} {\bibfnamefont {T.}~\bibnamefont
  {Prijon}}, \bibinfo {author} {\bibfnamefont {S.~D.}\ \bibnamefont {Huber}},\
  and\ \bibinfo {author} {\bibfnamefont {K.-E.}\ \bibnamefont {Huhtinen}},\
  }\bibfield  {title} {\bibinfo {title} {Data for "superfluid stiffness of
  superconductors with delicate topology"},\ }\href
  {https://doi.org/10.5281/zenodo.17975566} {10.5281/zenodo.17975566} (\bibinfo
  {year} {2026})\BibitemShut {NoStop}%
\end{thebibliography}

\begin{thebibliography}{1}%
\makeatletter
\providecommand \@ifxundefined [1]{%
 \@ifx{#1\undefined}
}%
\providecommand \@ifnum [1]{%
 \ifnum #1\expandafter \@firstoftwo
 \else \expandafter \@secondoftwo
 \fi
}%
\providecommand \@ifx [1]{%
 \ifx #1\expandafter \@firstoftwo
 \else \expandafter \@secondoftwo
 \fi
}%
\providecommand \natexlab [1]{#1}%
\providecommand \enquote  [1]{``#1''}%
\providecommand \bibnamefont  [1]{#1}%
\providecommand \bibfnamefont [1]{#1}%
\providecommand \citenamefont [1]{#1}%
\providecommand \href@noop [0]{\@secondoftwo}%
\providecommand \href [0]{\begingroup \@sanitize@url \@href}%
\providecommand \@href[1]{\@@startlink{#1}\@@href}%
\providecommand \@@href[1]{\endgroup#1\@@endlink}%
\providecommand \@sanitize@url [0]{\catcode `\\12\catcode `\$12\catcode
  `\&12\catcode `\#12\catcode `\^12\catcode `\_12\catcode `\%12\relax}%
\providecommand \@@startlink[1]{}%
\providecommand \@@endlink[0]{}%
\providecommand \url  [0]{\begingroup\@sanitize@url \@url }%
\providecommand \@url [1]{\endgroup\@href {#1}{\urlprefix }}%
\providecommand \urlprefix  [0]{URL }%
\providecommand \Eprint [0]{\href }%
\providecommand \doibase [0]{https://doi.org/}%
\providecommand \selectlanguage [0]{\@gobble}%
\providecommand \bibinfo  [0]{\@secondoftwo}%
\providecommand \bibfield  [0]{\@secondoftwo}%
\providecommand \translation [1]{[#1]}%
\providecommand \BibitemOpen [0]{}%
\providecommand \bibitemStop [0]{}%
\providecommand \bibitemNoStop [0]{.\EOS\space}%
\providecommand \EOS [0]{\spacefactor3000\relax}%
\providecommand \BibitemShut  [1]{\csname bibitem#1\endcsname}%
\let\auto@bib@innerbib\@empty
\bibitem [{\citenamefont {Chen}\ \emph {et~al.}(2024)\citenamefont {Chen},
  \citenamefont {Lin},\ and\ \citenamefont {Kao}}]{Chen2024}%
  \BibitemOpen
  \bibfield  {author} {\bibinfo {author} {\bibfnamefont {Y.-C.}\ \bibnamefont
  {Chen}}, \bibinfo {author} {\bibfnamefont {Y.-P.}\ \bibnamefont {Lin}},\ and\
  \bibinfo {author} {\bibfnamefont {Y.-J.}\ \bibnamefont {Kao}},\ }\bibfield
  {title} {\bibinfo {title} {Chern dartboard insulator: sub-Brillouin zone
  topology and skyrmion multipoles},\ }\href
  {https://doi.org/10.1038/s42005-023-01502-8} {\bibfield  {journal} {\bibinfo
  {journal} {Commun. Phys.}\ }\textbf {\bibinfo {volume} {7}},\ \bibinfo
  {pages} {32} (\bibinfo {year} {2024})}\BibitemShut {NoStop}%
\end{thebibliography}
\end{document}